\documentclass[journal]{IEEEtran}
\IEEEoverridecommandlockouts

\usepackage{amsmath,amsfonts}
\usepackage{algorithm}
\usepackage{algorithmic}
\usepackage{array}
\usepackage[caption=false,font=normalsize,labelfont=sf,textfont=sf]{subfig}
\usepackage{textcomp}
\usepackage{stfloats}
\usepackage{url}
\usepackage{verbatim}
\usepackage{graphicx}
\usepackage{cite}
\usepackage{color}

\usepackage{amssymb}
\newcommand {\mymarginpar}[1]{\marginpar{#1}}
\renewcommand {\marginpar}[1]{}

\def\_{\rule{.3em}{.15ex}}      

\newcommand{\ls}[1]
   {\dimen0=\fontdimen6\the\font
    \lineskip=#1\dimen0
    \advance\lineskip.5\fontdimen5\the\font
    \advance\lineskip-\dimen0
    \lineskiplimit=.9\lineskip
    \baselineskip=\lineskip
    \advance\baselineskip\dimen0
    \normallineskip\lineskip
    \normallineskiplimit\lineskiplimit
    \normalbaselineskip\baselineskip
    \ignorespaces
   }

\newcommand {\bearn}{\begin{eqnarray*}}
\newcommand {\eearn}{\end{eqnarray*}}
\newcommand {\barr}{\begin{array}}
\newcommand {\earr}{\end{array}}

\newcommand {\N}{{\cal N}}

\newtheorem{definition}{Definition}
\newtheorem{property}[definition]{Property}
\newtheorem{proposition}[definition]{Proposition}
\newtheorem{lemma}[definition]{Lemma}
\newtheorem{theorem}[definition]{Theorem}
\newtheorem{corollary}[definition]{Corollary}
\newtheorem{example}[definition]{Example}
\newtheorem{remark}[definition]{Remark}

\newcommand {\benum} {\begin{enumerate}}
\newcommand {\eenum} {\end{enumerate}}

\newcommand {\bdesc} {\begin{description}}
\newcommand {\edesc} {\end{description}}

\newcommand {\bfig}[2] {\begin{figure}
  \centering
  \includegraphics[width=#2]{#1}}
\newcommand {\brotatefig}[2] {\begin{figure}[htbp]
                        \centerline {
                         \epsfig{figure={#1},clip=,angle=-90,width={#2}}}}
\newcommand {\bfigfirst}[2] {\begin{figure}[h]
                        \centerline {
                        \setlength{\epsfxsize}{#2}
                        \epsffile{#1}}}
\newcommand {\efig}[2]{ \caption{#2}
                        \label{fig:#1}
                        \end{figure}
                        \mymarginpar{fig:#1}}
\newcommand {\erotatefig}[2]{ \caption{#2}
                        \label{fig:#1}
                        \end{figure}
                        \mymarginpar{fig:#1}}

\newcommand {\btab}[1]{
                       \begin{table}
                       \centering
                       \begin{tabular}{#1}}
\newcommand {\etab}[3] {
                       \end{tabular}
                       \caption[#3]{#2}
                       \label{tab:#1}
                       \end{table}
                       \mymarginpar{tab:#1}
                       \vspace{.1in}}

\newcommand {\btabular}[1]{\begin{center}
                       \begin{tabular}{#1}}
\newcommand {\etabular}{\end{tabular}
                       \end{center}}

\newcommand {\bdefin}[1]{\begin{definition}
                      \mymarginpar{def:#1}
                      \label{def:#1} }
\newcommand {\edefin}       {\end{definition}}

\newcommand {\bpro}[1]{\begin{property}
                      \mymarginpar{pro:#1}
                      \label{pro:#1} }
\newcommand {\epro}   {\end{property}}

\newcommand {\bprop}[1]{\begin{proposition}
                      \mymarginpar{prop:#1}
                      \label{prop:#1} }
\newcommand {\eprop}       {\end{proposition}}

\newcommand {\blem}[1]{\begin{lemma}
                      \mymarginpar{lem:#1}
                      \label{lem:#1} }
\newcommand {\elem}   {\end{lemma}}

\newcommand {\bthe}[1]{\begin{theorem}
                      \mymarginpar{the:#1}
                      \label{the:#1} }
\newcommand {\ethe}   {\end{theorem}}
\newcommand {\rthe}[1]{Theorem \ref{the:#1}}

\newcommand {\bproof}{\noindent {\bf Proof.} \ }
\newcommand {\eproof} {\hfill \squares \\ \vspace{.2cm}}
\newcommand {\bcor}[1]{\begin{corollary}
                      \mymarginpar{cor:#1}
                      \label{cor:#1} }
\newcommand {\ecor}   {\end{corollary}}

\newcommand {\bax}[1]{\begin{axiom}
                      \mymarginpar{ax:#1}
                      \label{ax:#1} }
\newcommand {\eax}       {\vspace{-.1in} \end{axiom}}

\newcommand {\bex}[2]{\vspace{.1in}
                      \begin{example}
                      \mymarginpar{ex:#1}
                       {\bf #2}
                      \label{ex:#1} }
\newcommand {\eex}       {\end{example} \vspace{.3cm} }

\newcommand {\brem}[1]{\begin{remark}
                      \mymarginpar{rem:#1}
                      \label{rem:#1} \em }
\newcommand {\erem}   {\end{remark}}

\newcommand {\beq}[1]{\mymarginpar{eq:#1}
                      \begin{equation}
                      \label{eq:#1} }

\newcommand {\beqno}[1]{\mymarginpar{eq:#1}
                      \begin{eqnarray}
                      \nonumber}

\newcommand {\eeq}       {\end{equation}}
\newcommand {\eeqno}       { && \end{eqnarray}}
\newcommand {\req}[1]{(\ref{eq:#1})}

\newcommand {\bear}[1]{\mymarginpar{eq:#1}
                       \begin{eqnarray}
                       \label{eq:#1} }

\newcommand {\bearno}[1]{\mymarginpar{eq:#1}
                       \begin{eqnarray}
                       \nonumber}

\newcommand {\eear}{\end{eqnarray}}
\newcommand {\eearno}{\end{eqnarray}}
\newcommand {\bsel}{\left \{ \begin{array}{cl}}
\newcommand {\esel}{\end{array} \right.}

\newcommand {\bmat}[1]{\left [ \begin{array}{#1}}
\newcommand {\emat}{\end{array} \right ]}
\newcommand {\bsec}[2]{\mymarginpar{sec:#2}
                       \section{#1}
                       \label{sec:#2} }

\newcommand {\bsubsec}[2]{\mymarginpar{sec:#2}
                       \subsection{#1}
                       \label{sec:#2} }

\def\R{I\kern-0.30em R}
\def\N{I\kern-0.30em N}
\def\P{I\kern-0.30em P}
\newcommand\squares{\vrule height6pt width7pt depth1pt}

\begin{document}

\title{Dynamic Hierarchical Birkhoff-von Neumann Decomposition for All-to-All GPU Communication}

\author{Yen-Chieh~Wu,
Cheng-Shang~Chang,~\IEEEmembership{Fellow,~IEEE,}
Duan-Shin~Lee,~\IEEEmembership{Senior Member,~IEEE,}
and H.~Jonathan~Chao,~\IEEEmembership{Life Fellow,~IEEE}%
\thanks{Yen-Chieh Wu, Cheng-Shang Chang, and Duan-Shin Lee are with the Institute of Communications Engineering, National Tsing Hua University, Hsinchu 30013, Taiwan, R.O.C.
(e-mail: jaywu2002@gapp.nthu.edu.tw; cschang@ee.nthu.edu.tw; lds@cs.nthu.edu.tw). Corresponding author: Yen-Chieh Wu.}%
\thanks{H. Jonathan Chao is with the Department of Electrical and Computer Engineering, New York University, New York, NY, USA
(e-mail: chao@nyu.edu).}%
}

\maketitle
\pagestyle{empty}
\begin{abstract}
All-to-all GPU communication is a critical bottleneck in large-scale training
clusters, where completion time is constrained by per-port bandwidth and can be
severely impacted by traffic skew across GPUs and network interface cards
(NICs). This issue is amplified by the two-tier structure of modern GPU systems,
which combine fast intra-server links with much slower inter-server networks.
Motivated by recent system observations that highlight the importance of traffic
reshaping and hierarchy awareness, we study all-to-all scheduling from an online
switching and queueing-theoretic perspective.

We propose a dynamic hierarchical Birkhoff--von Neumann (BvN) decomposition
framework tailored to two-tier GPU fabrics. At each frame boundary, traffic is
first balanced within each server using simple local operations to mitigate
micro-level GPU/NIC skew while preserving aggregate server-to-server demand. A
hierarchical BvN decomposition is then applied at the server level and refined
into GPU-level matchings, significantly reducing decomposition complexity
relative to a flat GPU-level approach. By integrating this construction with the
dynamic frame sizing (DFS) principle, we obtain an online scheduler with
provable stability under admissible Poisson arrivals. Simulations demonstrate
substantial reductions in mean frame length, particularly under
server-localized hotspot traffic.
\end{abstract}


\bsec{Introduction}{introduction}

Large-scale machine learning training increasingly relies on multi-GPU,
multi-server clusters, where \emph{communication}, rather than computation,
often dominates the end-to-end iteration time. In particular, modern workloads
such as mixture-of-experts (MoE) models and large-scale recommendation systems
repeatedly invoke \emph{all-to-all} collectives (e.g., \texttt{alltoallv}), making
overall performance highly sensitive to traffic skew, per-port contention, and
load imbalance across GPUs and network interface cards (NICs)
\cite{lepikhin2020gshard,fedus2021switch}.

A key challenge stems from the inherently \emph{two-tier} structure of modern GPU
clusters. GPUs within the same server are interconnected by high-bandwidth
scale-up fabrics such as NVLink or NVSwitch, whereas inter-server communication
relies on significantly slower scale-out networks such as InfiniBand or
Ethernet. This disparity is substantial in current platforms—for example,
NVIDIA reports up to \(3.6~\mathrm{TB/s}\) bidirectional NVLink bandwidth per GPU,
whereas a ConnectX-9 SuperNIC supports up to \(800~\mathrm{Gb/s}\) per-port
networking—making the scale-out tier a persistent bottleneck under skewed
all-to-all traffic \cite{nvidia_nvlink_page,connectx9_user_manual}. As a result,
micro-level imbalance across GPUs or NICs can create stragglers that dominate
completion time and waste available bandwidth on the bottleneck tier.

Motivated by this phenomenon, recent systems work has emphasized the importance
of structure-aware scheduling and traffic reshaping for GPU clusters. Chronos
advocates prescheduled circuit switching to reduce contention during large
language model (LLM) training \cite{renganathan2025chronos}. FAST demonstrates
that reshaping traffic over fast intra-server links can mitigate skew before
scheduling the inter-server fabric \cite{lei2025fast}. Optimized collective
libraries such as NCCL provide highly tuned implementations of collectives in
practice \cite{nccl16}, while topology-aware synthesis frameworks such as TACCL
\cite{taccl23} and traffic-engineering-based formulations such as TE-CCL
\cite{liu2024teccl} further highlight the benefits of exploiting hierarchy and
structure in collective communication. Complementary system mechanisms, such as
compiler-driven overlap of communication with computation \cite{jiang2024lancet},
hierarchical MoE communication primitives \cite{tutel2022}, generic
communication schedulers \cite{bytescheduler19}, in-network acceleration
\cite{switchml21}, and multi-NIC assistance via intra-server relaying
\cite{ren2025fuselink}, collectively establish \emph{why} traffic reshaping and
hierarchy matter in practice. However, these approaches largely rely on offline
planning, heuristics, or implementation-level optimizations, and typically do
not provide an \emph{online}, analytically tractable scheduling framework with
explicit per-port constraints and queueing-theoretic guarantees.

In this paper, we adopt a complementary perspective grounded in classical
switching and queueing theory. We seek an \emph{online} scheduling framework for
all-to-all GPU communication that (i) explicitly respects per-port matching
constraints, (ii) exploits the server/GPU hierarchy inherent in two-tier
clusters, and (iii) provides provable stability guarantees under stochastic
traffic. To this end, we revisit the classical Birkhoff--von Neumann (BvN)
decomposition framework \cite{Birkhoff46,von53}, which represents a traffic
matrix as a sequence of conflict-free matchings. From a scheduling viewpoint,
each (sub)permutation matrix specifies a stage in which every sender and
receiver participates in at most one transfer, naturally capturing per-port
constraints. BvN-type scheduling has also been widely used in switching systems
and traffic matrix decomposition, including input-buffered crossbar scheduling
and guaranteed-rate services \cite{CCH99,CCH2000}.

Despite its appealing structure, directly applying BvN decomposition at the GPU
granularity in large clusters faces two fundamental limitations. First,
decomposing an \(mn \times mn\) traffic matrix (with \(n\) servers and \(m\) GPUs
per server) can be computationally expensive and difficult to implement online.
Second, micro-level traffic skew within a server pair—precisely the phenomenon
observed in practice—can significantly inflate the number of required matchings,
inducing stragglers on the scale-out tier even when server-level traffic is
balanced.

To overcome these limitations, we develop a \emph{dynamic hierarchical}
Birkhoff--von Neumann decomposition tailored to two-tier GPU fabrics. At each
frame boundary, we first exploit the high-bandwidth intra-server network to
\emph{balance} traffic across GPUs and NICs within each server, mitigating
micro-level skew while preserving the aggregate server-to-server demand. This
balancing step formalizes, in an analytically controlled manner, the traffic
reshaping intuition that underlies many recent system designs
\cite{lei2025fast,ren2025fuselink}. We then perform a \emph{hierarchical} BvN
decomposition that operates primarily at the server level and refines server-pair
matchings into GPU-level matchings, substantially reducing decomposition
complexity compared to a flat GPU-level approach.

Finally, to support stochastic arrivals and online operation, we integrate the
balanced hierarchical decomposition with the dynamic frame sizing (DFS)
principle \cite{Chang2009,Lien2013,chang2018greenput}. DFS has been successfully
used to stabilize crossbar switches and related systems under stochastic traffic
\cite{Chang2009,Lien2013}, and here it enables an online scheduler whose backlog
evolution decouples cleanly across frames. This integration yields
queueing-theoretic stability guarantees expressed in terms of server-level
aggregate loads, while retaining sensitivity to micro-level imbalance through
the balancing step.

In summary, this paper bridges the gap between system-level observations about
traffic skew in GPU clusters and a principled scheduling framework with rigorous
guarantees. The main contributions are as follows:
\begin{enumerate}
\item We develop a hierarchical BvN decomposition for \((m,n)\)-block traffic
matrices that exploits server-level structure to reduce decomposition
complexity.
\item We design a simple, constructive intra-server traffic balancing algorithm
that mitigates GPU/NIC stragglers while preserving aggregate server-to-server
demand.
\item We build a DFS-based online scheduler on top of the balanced hierarchical
decomposition and establish queue stability under admissible Poisson arrivals.
\item Through simulation, we show that intra-server balancing substantially
reduces the mean frame length, with particularly pronounced gains under
server-localized hotspots and non-uniform per-GPU traffic.
\end{enumerate}

This paper is organized as follows.
In Section~\ref{sec:system}, we introduce the two-tier GPU communication model,
formulate the all-to-all scheduling problem under per-port matching constraints,
and derive a fundamental completion-time lower bound.
Section~\ref{sec:hier} develops the hierarchical
Birkhoff--von Neumann decomposition for $(m,n)$-block traffic matrices,
which forms the structural core of our scheduling framework.
Section~\ref{sec:balancing} presents the intra-server traffic balancing
algorithm and establishes its correctness and termination properties.
In Section~\ref{sec:DFS}, we integrate the balanced hierarchical
decomposition with the dynamic frame sizing principle and provide
queueing-theoretic stability analysis under stochastic arrivals.
Section~\ref{sec:simulation1} reports simulation results under both uniform and
server-localized hotspot traffic models.
Finally, Section~\ref{sec:conclusion} concludes the paper and discusses
directions for future research.

\section{System Model}
\label{sec:system}

We consider an all-to-all GPU communication model as in
\cite{renganathan2025chronos,lei2025fast}.
There are $n$ servers, each equipped with $m$ GPUs and $m$ network interface cards (NICs),
for a total of $mn$ GPUs and $mn$ NICs.
Each GPU is connected to a dedicated NIC whose ingress and egress bandwidths are both $B_2$.

The $mn$ NICs are interconnected by an $mn\times mn$ nonblocking crossbar switch, which can realize
any permutation matrix without conflict.
We index GPUs by $(i,\ell)$, where $i\in\{1,\ldots,n\}$ denotes the server index and
$\ell\in\{1,\ldots,m\}$ denotes the GPU index within a server.
Let $T_{(i,\ell)\to(j,k)}$ denote the amount of data transmitted from GPU $(i,\ell)$ to GPU $(j,k)$.
We exclude self-traffic, i.e.,
$T_{(i,\ell)\to(i,\ell)}=0$ for all $(i,\ell)$.

The objective is to find a transmission schedule that minimizes the \emph{completion time}
$T^{\min}$, defined as the time required for all GPUs to complete their data transfers.
Define the total outgoing traffic from GPU $(i,\ell)$ as
\begin{equation}
T^{\rm out}_{(i,\ell)} \triangleq \sum_{j=1}^n \sum_{k=1}^m T_{(i,\ell)\to(j,k)},
\label{eq:out}
\end{equation}
and the total incoming traffic to GPU $(j,k)$ as
\begin{equation}
T^{\rm in}_{(j,k)} \triangleq \sum_{i=1}^n \sum_{\ell=1}^m T_{(i,\ell)\to(j,k)}.
\label{eq:in}
\end{equation}

Since each NIC can transmit and receive at most $B_2$ units of data per unit time,
the standard port-capacity argument yields the lower bound
\begin{equation}
T^{\min} \ge
\frac{1}{B_2}
\max\!\left\{
\max_{(i,\ell)} T^{\rm out}_{(i,\ell)},
\max_{(j,k)} T^{\rm in}_{(j,k)}
\right\}.
\label{eq:lb}
\end{equation}
It is well known that this bound is achievable via a Birkhoff--von Neumann decomposition
\cite{Birkhoff46,von53}, which has been widely applied in
SS/TDMA systems \cite{Inukai79},
ATM input-buffered switches \cite{HKM98},
Birkhoff--von Neumann switches \cite{CCH2000,CCH2001},
and hybrid circuit/packet networks \cite{hybrid15}.

\subsection{Two-Tier Fabric and Intra-Server Load Balancing}

\begin{figure}[t]
  \centering
  \includegraphics[width=0.95\linewidth]{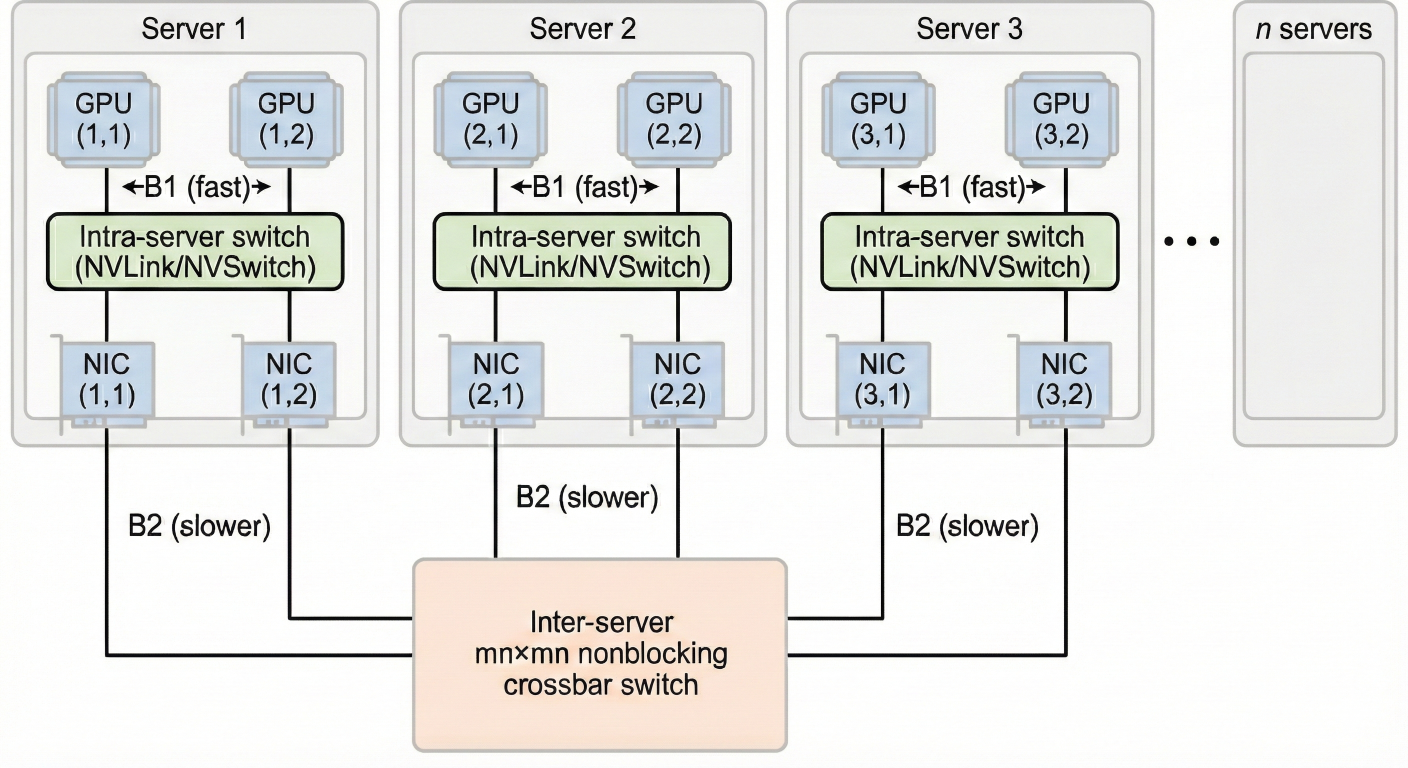}
  \caption{Two-tier GPU communication fabric. GPUs within the same server exchange traffic via a high-bandwidth intra-server switch (bandwidth $B_1$ per port), while inter-server GPU traffic is carried by NICs through the global $mn\times mn$ crossbar with bandwidth $B_2$ per port.}
  \label{fig:two_tier}
\end{figure}

Following \cite{lei2025fast}, we further assume that each server is equipped with a high-speed
intra-server switch connecting its $m$ GPUs.
Each port of this switch has bandwidth $B_1$, where $B_1 \gg B_2$.
This yields a heterogeneous two-tier fabric:
fast intra-server links (e.g., NVLink) and slower inter-server links (e.g., InfiniBand),
as illustrated in Fig.~\ref{fig:two_tier}.

Under this model, intra-server traffic does not consume NIC bandwidth.
Accordingly, the effective inter-server outgoing traffic from GPU $(i,\ell)$ is
\begin{equation}
T^{\rm out,inter}_{(i,\ell)} \triangleq
\sum_{j\neq i} \sum_{k=1}^m T_{(i,\ell)\to(j,k)},
\label{eq:out-inter}
\end{equation}
and the effective inter-server incoming traffic to GPU $(j,k)$ becomes
\begin{equation}
T^{\rm in,inter}_{(j,k)} \triangleq
\sum_{i\neq j} \sum_{\ell=1}^m T_{(i,\ell)\to(j,k)}.
\label{eq:in-inter}
\end{equation}

Hence, in the two-tier fabric, the lower bound in~\eqref{eq:lb} continues to
hold when the total outgoing and incoming loads,
$T^{\rm out}_{(i,\ell)}$ and $T^{\rm in}_{(j,k)}$, are replaced by their
effective inter-server counterparts defined in
\eqref{eq:out-inter}--\eqref{eq:in-inter}.

To further reduce the completion time, we are motivated by the key observation
in \cite{lei2025fast} that the high-bandwidth intra-server fabric can be used to
reshape traffic before it traverses the bottleneck inter-server tier.
In a two-tier cluster, stragglers often arise not from imbalance at the
\emph{server} level, but from micro-level skew across GPUs or NICs within the
same server. This suggests that intra-server load balancing can mitigate
per-port congestion without altering the aggregate server-to-server demand.

Suppose that the lower bound in \eqref{eq:lb} is dominated by
$T^{\rm out,inter}_{(i^*,\ell^*)}$, so that GPU $(i^*,\ell^*)$ is the system
straggler. Then
\begin{align}
T^{\rm out,inter}_{(i^*,\ell^*)}
&\ge \frac{1}{m}\sum_{\ell=1}^m T^{\rm out,inter}_{(i^*,\ell)} \nonumber\\
&= \sum_{j\neq i^*} \sum_{k=1}^m
\frac{1}{m}\sum_{\ell=1}^m T_{(i^*,\ell)\to(j,k)}.
\label{eq:avg-out}
\end{align}
The inequality shows that the maximum outgoing load at server $i^*$ is at least
the average outgoing load across its $m$ GPUs. Hence, if traffic from server
$i^*$ to remote GPUs is unevenly distributed across its local GPUs, the
straggler effect is attributable to micro-level imbalance rather than to
excess aggregate demand.

Exploiting the high-speed intra-server switch, traffic from server $i^*$ to any
remote GPU $(j,k)$ can be evenly redistributed across the $m$ GPUs within that
server, yielding the balanced traffic matrix
\begin{equation}
T'_{(i^*,\ell')\to(j,k)}
\triangleq
\frac{1}{m}\sum_{\ell=1}^m T_{(i^*,\ell)\to(j,k)},
\qquad \forall \;\ell'=1,\ldots,m.
\label{eq:balance-out}
\end{equation}
This redistribution preserves the total server-to-server traffic while
equalizing per-GPU outgoing loads. Consequently, it cannot increase the maximum
outgoing load and therefore cannot increase the completion time; it strictly
reduces the completion time unless the traffic is already evenly distributed
across the $m$ GPUs within server $i^*$.

An analogous argument applies when the dominant term in \eqref{eq:lb} is
$T^{\rm in,inter}_{(j^*,k^*)}$: by balancing incoming traffic within server
$j^*$ using intra-server transfers, one can similarly mitigate micro-level
imbalance on the receiving side without affecting aggregate demand.

\subsection{Server-Level Traffic Aggregation}
\label{sec:aggregate}

The preceding balancing arguments suggest that, in a two-tier fabric,
completion time is fundamentally governed by \emph{server-level} aggregate
traffic rather than by micro-level GPU imbalance. We therefore formalize this
aggregation explicitly.

For each ordered pair of servers $(i,j)$, define the aggregated traffic
\begin{equation}
T_{i,j}
\triangleq
\sum_{\ell=1}^m \sum_{k=1}^m
T_{(i,\ell)\to(j,k)},
\label{eq:agg}
\end{equation}
which represents the total amount of data transmitted from server $i$ to
server $j$ across all GPUs.

If both outgoing and incoming intra-server balancing are performed, then for
each server pair $(i,j)$ the traffic can be evenly distributed across the
$m \times m$ GPU pairs. The resulting fully balanced traffic matrix becomes
\begin{equation}
\widetilde{T}_{(i,\ell)\to(j,k)}
\triangleq
\frac{1}{m^2} T_{i,j},
\qquad \forall\, i,\ell,j,k.
\label{eq:full-balance}
\end{equation}
By construction, this transformation preserves the aggregate server-to-server
demand $T_{i,j}$ while equalizing both the outgoing load at each GPU in server
$i$ and the incoming load at each GPU in server $j$.

Applying the per-port capacity lower bound to the balanced traffic in
\eqref{eq:full-balance}, and noting that each GPU in server $i$ now carries an
outgoing load of $\sum_{j=1}^n T_{i,j}/m$ and each GPU in server $j$ carries
an incoming load of $\sum_{i=1}^n T_{i,j}/m$, we obtain
\begin{equation}
\widetilde{T}^{\min}
\ge
\frac{1}{B_2}
\max\!\left\{
\max_i \sum_{j=1}^n \frac{T_{i,j}}{m},
\max_j \sum_{i=1}^n \frac{T_{i,j}}{m}
\right\}.
\label{eq:balanced-lb}
\end{equation}
This bound depends only on server-level aggregates $\{T_{i,j}\}$, thereby
demonstrating that, after intra-server balancing, the completion time is
governed by server-level traffic rather than by micro-level GPU skew.

\section{Hierarchical Birkhoff--von Neumann Decomposition}
\label{sec:hier}

The lower bound in \req{balanced-lb} for the aggregated inter-server traffic
(see Section~\ref{sec:aggregate}) can be achieved by applying a
Birkhoff--von Neumann (BvN) decomposition to the $mn\times mn$ traffic matrix
among all NICs.
However, directly decomposing such a large matrix incurs prohibitive
computational complexity.
We therefore propose a \emph{hierarchical} Birkhoff--von Neumann decomposition,
which exploits the server structure to significantly reduce complexity.

\subsection{Preliminaries}

We begin by introducing basic matrix notions used throughout the paper.

\paragraph{Permutation and subpermutation matrices}
A \emph{permutation matrix} is a binary-valued square matrix in which each row and
each column contains exactly one entry equal to~1.
A \emph{subpermutation matrix} is a binary-valued square matrix in which each row and
each column contains at most one entry equal to~1.
Equivalently, a permutation (resp.\ subpermutation) matrix represents a perfect
(resp.\ partial) matching in a bipartite graph.

\begin{definition}[Scaled doubly stochastic matrices]
An $n\times n$ nonnegative matrix $\mathbf{X}$ is called \emph{doubly stochastic with scale}
$\Delta$ if all its row sums and all its column sums are equal to~$\Delta$.
An $n\times n$ nonnegative matrix $\mathbf{X}$ is called \emph{doubly substochastic with scale}
$\Delta$ if all its row sums and all its column sums are not greater than~$\Delta$.
\end{definition}

The following classical result, due to Birkhoff \cite{Birkhoff46} and von Neumann \cite{von53}, characterizes the
structure of such matrices when their entries are integers.

\begin{proposition}[Birkhoff--von Neumann decomposition]
\label{prop:bvn}
Let $\mathbf{X}$ be an $n\times n$ nonnegative integer-valued matrix.
\begin{enumerate}
\item If $\mathbf{X}$ is doubly stochastic with scale $\Delta$, then $\mathbf{X}$ can be
decomposed as
\[
\mathbf{X}=\sum_{d=1}^{\Delta} \mathbf{P}_d,
\]
where each $\mathbf{P}_d$ is an $n\times n$ permutation matrix.
\item If $\mathbf{X}$ is doubly substochastic with scale $\Delta$, then $\mathbf{X}$ can be
decomposed as
\[
\mathbf{X}=\sum_{d=1}^{D} \mathbf{Q}_d,
\]
where $D\le \Delta$ and each $\mathbf{Q}_d$ is an $n\times n$ subpermutation matrix.
\end{enumerate}
\end{proposition}

\bproof
We outline a self-contained proof based on bipartite graph edge coloring.

Consider the bipartite multigraph $G=(\mathcal{U}\cup\mathcal{V},E)$ with
$\mathcal{U}=\{1,\ldots,n\}$ and $\mathcal{V}=\{1,\ldots,n\}$, where the number of parallel
edges between $u\in\mathcal{U}$ and $v\in\mathcal{V}$ equals $X_{u,v}$.
Each row sum and column sum of $\mathbf{X}$ equals the degree of the corresponding vertex.

\emph{Doubly stochastic case.}
If $\mathbf{X}$ is doubly stochastic with scale $\Delta$, then $G$ is a $\Delta$-regular
bipartite multigraph. By K\H{o}nig's line-coloring theorem, the edges of $G$ can be colored
using exactly $\Delta$ colors such that no two edges of the same color share an endpoint.
Each color class therefore forms a perfect matching in $G$, which corresponds to a
permutation matrix. Summing these $\Delta$ permutation matrices yields $\mathbf{X}$.

\emph{Doubly substochastic case.}
If $\mathbf{X}$ is doubly substochastic with scale $\Delta$, then the maximum degree of $G$
is at most $\Delta$. By adding dummy edges to $G$ if necessary, we can obtain a
$\Delta$-regular bipartite multigraph without affecting the original edges.
Applying the above argument to the augmented graph and discarding the dummy edges yields
a decomposition of $\mathbf{X}$ into at most $\Delta$ subpermutation matrices.
\eproof

\bsubsec{The Theory of Hierarchical Birkhoff--von Neumann Decomposition}{theory}

In this section, we develop the main theory for the hierarchical Birkhoff--von Neumann decomposition.

\begin{definition}[$(m,n)$-block matrix]
\label{def:block}
An $mn\times mn$ matrix $\mathbf{X}$ is called an \emph{$(m,n)$-block matrix} if it can be partitioned
into $n\times n$ blocks, where each block is an $m\times m$ submatrix. That is,
\[
\mathbf{X}
=
\begin{bmatrix}
\mathbf{X}_{1,1} & \mathbf{X}_{1,2} & \cdots & \mathbf{X}_{1,n} \\
\mathbf{X}_{2,1} & \mathbf{X}_{2,2} & \cdots & \mathbf{X}_{2,n} \\
\vdots & \vdots & \ddots & \vdots \\
\mathbf{X}_{n,1} & \mathbf{X}_{n,2} & \cdots & \mathbf{X}_{n,n}
\end{bmatrix},
\]
where each block $\mathbf{X}_{i,j}\in\mathbb{R}^{m\times m}$.
Equivalently, each block row and each block column of $\mathbf{X}$ consists of exactly $n$
$m\times m$ submatrices.
\end{definition}

\bthe{hierBvN} Consider an \emph{$(m,n)$-block matrix} $\mathbf{X}$ in Definition \ref{def:block}.
Assume that $\mathbf{X}_{i,j}$ is an integer-valued doubly substochastic matrix with scale $\Delta_{i,j}$.
Define 
\beq{hier1111}\Delta= \max[ \max_{1\le i \le n}\sum_{j=1}^n \Delta_{i,j}, \max_{1 \le j \le n} \sum_{i=1}^n \Delta_{i,j}].
\eeq
Then there are $\Delta$ $mn \times mn$ subpermutation matrices $\{P_{d}, 1\le d \le \Delta\}$
such that 
\beq{hier2222} 
\mathbf{X}=\sum_{d=1}^\Delta  P_{d}.
\eeq
\ethe

We prove the theorem by constructing the subpermutation matrices $\{P_{d}\}$ explicitly.

\medskip
\noindent{\bf Step 1: Decompose each block into subpermutation matrices.}
Fix any $(i,j)\in\{1,\ldots,n\}^2$. By assumption, the block $\mathbf{X}_{i,j}$ is an integer-valued
doubly substochastic matrix with scale $\Delta_{i,j}$. By Proposition~\ref{prop:bvn}, there exist
subpermutation matrices $\mathbf{S}^{(i,j)}_{1},\ldots,\mathbf{S}^{(i,j)}_{\Delta_{i,j}} \in \{0,1\}^{m\times m}$
such that
\begin{equation}
\mathbf{X}_{i,j}=\sum_{r=1}^{\Delta_{i,j}} \mathbf{S}^{(i,j)}_{r}.
\label{eq:block-decomp}
\end{equation}
(If fewer than $\Delta_{i,j}$ subpermutation matrices are returned by Proposition~\ref{prop:bvn},
we append all-zero $m\times m$ matrices so that \eqref{eq:block-decomp} holds with exactly $\Delta_{i,j}$ terms.)

\medskip
\noindent{\bf Step 2: Construct a server-level matrix of scales.}
Define the $n\times n$ nonnegative integer matrix
\[
A \triangleq [A_{i,j}],\qquad A_{i,j}\triangleq \Delta_{i,j}.
\]
By definition of $\Delta$ in \req{hier1111}, the maximum row sum and maximum column sum of $A$ are
both at most $\Delta$. Hence $A$ is an integer-valued doubly substochastic matrix with scale $\Delta$.

By Proposition~\ref{prop:bvn}, there exist subpermutation matrices
$Q_1,\ldots,Q_{\Delta}\in\{0,1\}^{n\times n}$ such that
\begin{equation}
A=\sum_{d=1}^{\Delta} Q_d .
\label{eq:A-decomp}
\end{equation}
(Again, if fewer than $\Delta$ terms arise, pad with all-zero $n\times n$ matrices.)

\medskip
\noindent{\bf Step 3: Allocate each block-level component to one global index $d$.}
For each fixed block $(i,j)$, equation \eqref{eq:A-decomp} implies
\beq{hier3333}
\Delta_{i,j}=A_{i,j}=\sum_{d=1}^{\Delta} (Q_d)_{i,j}.
\eeq
Let
\[
\mathcal{D}_{i,j}\triangleq \{ d : (Q_d)_{i,j}=1\}.
\]
From \req{hier3333}, we know that
$|\mathcal{D}_{i,j}|=\Delta_{i,j}$.
Arrange the $\Delta_{i,j}$ elements in $\mathcal{D}_{i,j}$ in increasing order and we can represent the
set 
$$\mathcal{D}_{i,j}=\{d(i,j,r), r=1,2, \ldots, \Delta_{i,j}\}.$$
Since $\mathcal{D}_{i,j}$ is a set with $\Delta_{i,j}$ distinct elements, the mapping
$r\mapsto d(i,j,r)$ is a bijection from $\{1,\ldots,\Delta_{i,j}\}$ onto $\mathcal{D}_{i,j}$; in particular,
for each $d\in\mathcal{D}_{i,j}$ there exists a unique $r$ such that $d(i,j,r)=d$.

\medskip
\noindent{\bf Step 4: Build the global $mn\times mn$ subpermutation matrices $P_{d}$.}
For each $d\in\{1,\ldots,\Delta\}$, define an $mn\times mn$ $(m,n)$-block matrix
$P_{d}$ by specifying its $(i,j)$ block as follows:
\begin{equation}
(P_{d})_{i,j} \triangleq
\begin{cases}
\mathbf{S}^{(i,j)}_{r} & \text{if $d(i,j,r)=d$},\\[2mm]
\mathbf{0}_{m\times m} & \text{otherwise}.
\end{cases}
\label{eq:Pdr-def}
\end{equation}

\medskip
\noindent{\bf Step 5: Each $P_{d}$ is a subpermutation matrix.}
We verify that $P_{d}$ has at most one ``1'' in each row and each column.

Fix $d$. Because $Q_d$ is a subpermutation matrix, in each block row $i$ there is at most one block column $j$
with $(Q_d)_{i,j}=1$. Hence, in block row $i$ of $P_{d}$,
there is at most one nonzero $m\times m$ block. Inside that block, $(P_{d})_{i,j}$ is a subpermutation matrix,
so each of its $m$ rows contains at most one ``1''. Therefore, every row of the full $mn\times mn$ matrix $P_{d}$
contains at most one ``1''. The same argument applies to columns because each column of $Q_d$ contains at most one ``1'',
so each block column of $P_{d}$ has at most one nonzero block, and that block has at most one ``1'' per column.
Thus $P_{d}$ is a subpermutation matrix.

\medskip
\noindent{\bf Step 6: Summing $\{P_{d}\}$ recovers $\mathbf{X}$.}

Note that 
\[
\sum_{d=1}^{\Delta} (P_{d})_{i,j}
=
\sum_{r=1}^{\Delta_{i,j}} \mathbf{S}^{(i,j)}_{r}
=
\mathbf{X}_{i,j},
\]
where the last equality follows from \eqref{eq:block-decomp}.
Since this holds for every $(i,j)$ block, we conclude that
\[
\mathbf{X}=\sum_{d=1}^{\Delta} P_{d},
\]
which is exactly \req{hier2222}. This completes the proof of \rthe{hierBvN}.

\subsection{Using Hierarchical Birkhoff-von Neumann Decomposition for the Two-Tier Fabric}

In this section, we illustrate how to apply \rthe{hierBvN} to the balanced
traffic matrix in \eqref{eq:full-balanceb}, i.e.,
\begin{equation}
\widetilde{T}_{(i,\ell)\to(j,k)}
\triangleq
\frac{1}{m^2} T_{i,j},
\qquad \forall\, i,\ell,j,k,
\label{eq:full-balanceb}
\end{equation}
where $T_{i,j}$ is the aggregated traffic from server $i$ to server $j$ defined in \eqref{eq:agg}.
Assume that $T_{i,j}/m^2$ are nonnegative integers. 
Then $\widetilde{T}$ is an $(m,n)$-block matrix whose $(i,j)$ block equals
\begin{equation}
\widetilde{T}_{i,j} = \frac{T_{i,j}}{m^2}\,J_m,
\label{eq:block-form}
\end{equation}
where $J_m$ denotes the $m\times m$ all-ones matrix.

We illustrate Step 1 to Step 4 for the case with $m=2$, $n=3$. Specifically,
we consider $n=3$ servers, each with $m=2$ NICs (equivalently, $mn=6$ NICs total).
We enforce \emph{zero intra-server traffic}, i.e., the $(i,i)$ block is the $2\times 2$ zero matrix.
Suppose that ${T_{i,j}}/{m^2}=1$ for  $(i, j)=(1,2),(2,3)$ and $(3,1)$. Then
the corresponding $6\times 6$ balanced NIC-level matrix $\widetilde{T}$ in
\eqref{eq:full-balanceb} is an $(2,3)$-block matrix with $3\times 3$ blocks of size $2\times 2$:

\[
\widetilde{T}
=
\begin{bmatrix}
\mathbf{0} & \mathbf{1}\mathbf{1}^\top & \mathbf{0}\\
\mathbf{0} & \mathbf{0} & \mathbf{1}\mathbf{1}^\top\\
\mathbf{1}\mathbf{1}^\top & \mathbf{0} & \mathbf{0}
\end{bmatrix},
\mathbf{1}\mathbf{1}^\top=
\begin{bmatrix}
1&1\\
1&1
\end{bmatrix},
\mathbf{0}=
\begin{bmatrix}
0&0\\
0&0
\end{bmatrix}.
\]

\medskip
\noindent\textbf{Step 1: Decompose each block into subpermutation matrices.}
For each nonzero $2\times 2$ block $\mathbf{1}\mathbf{1}^\top$, we use the standard decomposition
\[
\mathbf{1}\mathbf{1}^\top
=
\underbrace{\begin{bmatrix}1&0\\0&1\end{bmatrix}}_{\mathbf{S}^{(i,j)}_{1}}
+
\underbrace{\begin{bmatrix}0&1\\1&0\end{bmatrix}}_{\mathbf{S}^{(i,j)}_{2}},
\]
where each $\mathbf{S}^{(i,j)}_{r}$ is a $2\times 2$ permutation matrix (hence a subpermutation matrix).
For a zero block, there is nothing to decompose.

Thus, for every inter-server pair $(i, j)=(1,2),(2,3)$ and $(3,1)$,
\[
\widetilde{T}_{i,j}=\mathbf{S}^{(i,j)}_{1}+\mathbf{S}^{(i,j)}_{2},
\qquad
\Delta_{i,j}=2,
\]
and $\Delta_{i,j}=0$ otherwise.

\medskip
\noindent\textbf{Step 2: Construct a server-level matrix of scales.}
The block-scale matrix is
\[
A \triangleq [A_{i,j}],\qquad A_{i,j}\triangleq \Delta_{i,j}.
\]
Thus,
\[
A\triangleq[\Delta_{i,j}]_{i,j=1}^3
=
\begin{bmatrix}
0 & 2 & 0\\
0 & 0 & 2\\
2 & 0 & 0
\end{bmatrix}.
\]
Its row sums and column sums are all equal to $2$, hence (in the terminology of the preliminaries)
$A$ is doubly stochastic with scale
\[
\Delta=\max\Big\{\max_i\sum_{j=1}^3 \Delta_{i,j},\ \max_j\sum_{i=1}^3 \Delta_{i,j}\Big\}=2.
\]

\medskip
\noindent\textbf{Step 3: Allocate each block-level component to one global index $d$.}
By Proposition~\ref{prop:bvn}, $A$ can be decomposed into $\Delta=2$ permutation matrices.
In this example,
\[
Q_1=Q_2=
\begin{bmatrix}
0 & 1 & 0\\
0 & 0 & 1\\
1 & 0 & 0
\end{bmatrix}.
\]
Hence, for each nonzero block location $(i,j)\in\{(1,2),(2,3),(3,1)\}$,
\[
\mathcal{D}_{i,j}\triangleq\{d\in\{1,2\}:(Q_d)_{i,j}=1\}=\{1,2\},
\]
and thus
\[
|\mathcal{D}_{i,j}|=\Delta_{i,j}=2.
\]

\medskip
\noindent\textbf{Step 4: Build the global $mn\times mn$ subpermutation matrices $P_{d}$.}
We index the $mn=6$ NICs in the order
\[
(1,1),(1,2),(2,1),(2,2),(3,1),(3,2).
\]
We now construct $P_{d}$ for $d\in\{1,2\}$.

\smallskip
\emph{(i) Matrix $P_{1}$:} place $\mathbf{S}^{(i,j)}_{1}=I_2$ in blocks $(1,2)$, $(2,3)$, $(3,1)$, and zeros elsewhere:
\[
P_{1}=
\begin{bmatrix}
0&0&1&0&0&0\\
0&0&0&1&0&0\\
0&0&0&0&1&0\\
0&0&0&0&0&1\\
1&0&0&0&0&0\\
0&1&0&0&0&0
\end{bmatrix}.
\]
This is a $6\times 6$ permutation matrix (hence a subpermutation matrix).

\smallskip
\emph{(ii) Matrix $P_{2}$:} place $\mathbf{S}^{(i,j)}_{2}=\begin{bmatrix}0&1\\1&0\end{bmatrix}$ in the same three blocks:
\[
P_{2}=
\begin{bmatrix}
0&0&0&1&0&0\\
0&0&1&0&0&0\\
0&0&0&0&0&1\\
0&0&0&0&1&0\\
0&1&0&0&0&0\\
1&0&0&0&0&0
\end{bmatrix}.
\]
This is also a $6\times 6$ permutation matrix.

\bsec{Balancing the Traffic within a Server}{balancing}

In this section, we present a simple constructive algorithm for balancing
traffic among the $m$ GPUs (or NICs) within a single server.
For the hierarchical Birkhoff--von Neumann decomposition developed in the
previous section, it is not necessary to achieve the fully balanced form
in~\req{full-balance}.  Instead, it suffices to ensure that each row sum
and each column sum of the corresponding $m\times m$ block is properly
balanced (i.e., bounded by the same target value).
The proposed algorithm operates on an $m\times m$ nonnegative integer matrix,
uses only local unit-transfer operations, preserves integrality and
nonnegativity, and is guaranteed to terminate in finite time.

\subsection{The objective of traffic balancing}

Recall that $T_{(i,\ell) \to (j,k)}$ denotes the amount of data  to be transmitted from the $(i,\ell)$ GPU to the $(j,k)$ GPU. Let
$$x_{\ell k} \triangleq T_{(i,\ell) \to (j,k)}$$
and form the $m \times m$ matrix 
 $\mathbf{X}=[x_{\ell k}]$ (here we omit the subscripts $i$ and $j$ for notational clarity).
Define
\beq{bal1111}
W \triangleq \sum_{\ell=1}^{m}\sum_{k=1}^{m} x_{\ell k},\qquad
r_\ell \triangleq \sum_{k=1}^{m} x_{\ell k},\qquad
c_k \triangleq \sum_{\ell=1}^{m} x_{\ell k},
\eeq
and let
\beq{bal2222}
B \triangleq \left\lceil \frac{W}{m}\right\rceil .
\eeq
Our objective is to transform $\mathbf{X}$ into another nonnegative
integer-valued $m\times m$ matrix, using local operations that preserve the total
sum $W$, such that
\[
r_\ell \le B \quad \forall\, \ell,\qquad
c_k \le B \quad \forall\, k .
\]
After the transform, the matrix is doubly substochastic with the scale $B$.

\subsection{Local Unit-Transfer Operations}

We employ two types of unit-transfer operations. Each operation preserves
nonnegativity, integrality, and the total sum $W$.

\subsubsection{Column Transfer}
For a fixed column $k$ and two distinct rows $\ell\neq \ell'$ with
$x_{\ell k}\ge 1$, perform
\[
x_{\ell k}\leftarrow x_{\ell k}-1,\qquad
x_{\ell' k}\leftarrow x_{\ell' k}+1 .
\]
This operation decreases the row sum $r_\ell$ by one and increases $r_{\ell'}$
by one, while leaving all column sums unchanged.

\subsubsection{Row Transfer}
For a fixed row $\ell$ and two distinct columns $k\neq k'$ with
$x_{\ell k}\ge 1$, perform
\[
x_{\ell k}\leftarrow x_{\ell k}-1,\qquad
x_{\ell k'}\leftarrow x_{\ell k'}+1 .
\]
This operation decreases the column sum $c_k$ by one and increases $c_{k'}$ by
one, while leaving all row sums unchanged.

\subsection{Two-Phase Balancing Algorithm}

\subsubsection{Phase~I: Row Balancing}

While there exists a row index $\ell$ such that $r_\ell>B$, select another row
$\ell'$ with $r_{\ell'}<B$ (such a row must exist since
$\sum_{\ell=1}^{m} r_\ell=W\le mB$), choose any column $k$ with $x_{\ell k}\ge 1$,
and apply a column transfer from $(\ell,k)$ to $(\ell',k)$.

Define the row imbalance potential
\[
\Phi_{\mathrm{row}} \triangleq \sum_{\ell=1}^{m} \max(0,r_\ell-B).
\]
Each column transfer reduces $\Phi_{\mathrm{row}}$ by exactly one, guaranteeing
finite termination with
\[
r_\ell \le B \quad \forall\, \ell .
\]

\subsubsection{Phase~II: Column Balancing}

After Phase~I, all row sums are fixed and satisfy $r_\ell\le B$.
While there exists a column index $k$ such that $c_k>B$, select another column
$k'$ with $c_{k'}<B$, choose any row $\ell$ with $x_{\ell k}\ge 1$, and apply a
row transfer from $(\ell,k)$ to $(\ell,k')$.

Define the column imbalance potential
\[
\Phi_{\mathrm{col}} \triangleq \sum_{k=1}^{m} \max(0,c_k-B).
\]
Each row transfer reduces $\Phi_{\mathrm{col}}$ by exactly one, ensuring finite
termination with
\[
c_k \le B \quad \forall\, k .
\]

\bthe{balancing}
Let $\mathbf{X}=[x_{\ell k}]\in\mathbb{Z}_{\ge 0}^{m\times m}$ be the initial
intra-server traffic matrix, and let $W$, $\{r_\ell\}$, $\{c_k\}$ and $B$ be defined as
in \req{bal1111} and \req{bal2222}. Then the two-phase balancing algorithm described above
satisfies the following properties:
\begin{enumerate}
  \item All matrix entries remain nonnegative integers throughout the execution
        of the algorithm.
  \item The total traffic volume $W$ is invariant.
  \item The algorithm terminates after at most
        $$\sum_{\ell=1}^{m} \max(0,r_\ell-B)+\sum_{k=1}^{m} \max(0,c_k-B)$$ 
        unit-transfer operations.
  \item Upon termination, the resulting matrix satisfies
        \[
        r_\ell \le \left\lceil \frac{W}{m}\right\rceil
        \quad \forall\, \ell\in\{1,\ldots,m\},
       \]
       and
       \[
        c_k \le \left\lceil \frac{W}{m}\right\rceil
        \quad \forall\, k\in\{1,\ldots,m\}.
        \]
\end{enumerate}
\ethe

\bproof
We prove each claim in turn.

\emph{(i) Preservation of nonnegativity and integrality.}
Each unit-transfer operation decreases one matrix entry by one and increases
another entry by one. Transfers are applied only when the decremented entry is at
least one, hence all entries remain nonnegative. Since all updates are integer
increments or decrements, integrality is preserved.

\emph{(ii) Invariance of the total traffic volume.}
Each unit transfer moves one unit of traffic from one entry to another. Therefore,
the sum of all matrix entries remains unchanged, and the total volume $W$ is
invariant throughout the algorithm.

\emph{(iii) Finite termination.}
During Phase~I, each column transfer reduces the row-imbalance potential
$\Phi_{\mathrm{row}}$ by exactly one, and $\Phi_{\mathrm{row}}$ is a nonnegative
integer-valued function. Hence Phase~I terminates after at most
$\Phi_{\mathrm{row}}(0)$ operations, where 
$$\Phi_{\mathrm{row}}(0)=\sum_{\ell=1}^{m} \max(0,r_\ell-B).$$
is the initial row-imbalance potential.
After Phase~I, all row sums satisfy $r_\ell\le \lceil W/m\rceil$.
During Phase~II, each row transfer reduces the column-imbalance potential
$\Phi_{\mathrm{col}}$ by exactly one, and $\Phi_{\mathrm{col}}$ is also a
nonnegative integer-valued function. Thus Phase~II terminates after at most
$\Phi_{\mathrm{col}}(0)$ operations, where 
$$\Phi_{\mathrm{col}}(0)=\sum_{k=1}^{m} \max(0,c_k-B)$$
is the initial column-imbalance potential.
The total number of unit-transfer operations is therefore at most
$\Phi_{\mathrm{row}}(0)+\Phi_{\mathrm{col}}(0)$.

\emph{(iv) Feasibility of the final matrix.}
By construction, Phase~I terminates only when $r_\ell\le B=\lceil W/m\rceil$ for
all rows~$\ell$. Phase~II leaves all row sums unchanged and terminates only when
$c_k\le B$ for all columns~$k$. Therefore, upon termination, the resulting matrix
satisfies the desired row and column sum bounds.
\eproof

\bsec{Dynamic Hierarchical Birkhoff--von Neumann Decomposition}{DFS}

In this section, we present a dynamic frame sizing (DFS) algorithm
\cite{Chang2009,Lien2013,chang2018greenput} built upon the
\emph{hierarchical Birkhoff--von Neumann (BvN) decomposition} developed in
Section~\ref{sec:hier}.

Throughout this section, we adopt a discrete-time model.
All packets are assumed to have identical size, and time is slotted so that,
after normalizing the inter-server bandwidth $B_2$, at most one packet can be
transmitted over a link in each time slot.
All switches are assumed to be empty at time~$0$, and packet arrivals occur
from time~$1$ onward.

\subsection{Dynamic Frame Sizing Principle}
\label{sec:steps}

The DFS algorithm operates in a frame-based manner and consists of the following steps:
\begin{description}
\item[(i)] Time is divided into frames. 
Packets arriving during a frame are buffered and are served only in the subsequent frame.
\item[(ii)] The frame length is adaptive rather than fixed. At the beginning of each frame,
the frame size is set to the minimum completion (clearance) time required to serve all packets
that have accumulated up to that time. If all buffers are empty, the frame length is set to~1.
\item[(iii)] Traffic is first balanced within each server using the intra-server balancing
procedure described in Section~\ref{sec:balancing}.
\item[(iv)] The hierarchical Birkhoff--von Neumann decomposition described in
Section~\ref{sec:hier} is then applied to determine the subpermutation matrices
used by the global $mn\times mn$ crossbar switch during the frame.
\end{description}

\subsection{Frame-Based Operation}

We consider an $mn\times mn$ input-buffered crossbar switch interconnecting all
GPUs.
At each $(i,\ell)$-th GPU, there are $mn$ virtual output queues (VOQs), one for
each destination $(j,k)$ GPU.
Packets arriving at the $(i,\ell)$ GPU and destined for the $(j,k)$ GPU are placed
in the corresponding VOQ.

Let $T_f$ denote the length of the $f$-th frame and define
\[
\tau_f \triangleq \sum_{r=1}^{f-1} T_r .
\]
Accordingly, $\tau_f+1$ and $\tau_{f+1}$ are the first and last time slots of the
$f$-th frame, respectively.

Let $P(t)=\big(P_{(i,\ell),(j,k)}(t)\big)$ denote the permutation matrix specifying
the connection pattern of the crossbar switch at time~$t$, where
$P_{(i,\ell),(j,k)}(t)=1$ if and only if the $(i,\ell)$ input is connected to the
$(j,k)$ output at time~$t$.
Let $a_{(i,\ell),(j,k)}(t)$ denote the number of packets arriving at time~$t$ to
the $(i,\ell)$ GPU that are destined for the $(j,k)$ GPU, and let
$x_{(i,\ell),(j,k)}(t)$ denote the number of packets in the corresponding VOQ at
the end of time slot~$t$.

Define
\[
A_{(i,\ell),(j,k)}(s,t)
\triangleq
\sum_{u=s+1}^{t} a_{(i,\ell),(j,k)}(u),
\]
which represents the number of packets arriving in the interval $(s,t]$ for the
traffic from the $(i,\ell)$ GPU to the $(j,k)$ GPU.

By definition of the minimum completion time and by applying the framed
hierarchical Birkhoff--von Neumann decomposition within each frame, the
permutation matrices $\{P(t)\}$ can be chosen such that all packets carried over
from the previous frame are cleared by the end of the current frame, i.e.,
\[
x_{(i,\ell),(j,k)}(\tau_f)
\le
\sum_{t=\tau_f+1}^{\tau_{f+1}} P_{(i,\ell),(j,k)}(t).
\]

As a consequence, the backlog at the beginning of the $(f+1)$-th frame consists
exactly of the packets that arrived during the $f$-th frame, namely,
\beq{gov3333}
x_{(i,\ell),(j,k)}(\tau_{f+1})
=
A_{(i,\ell),(j,k)}(\tau_f,\tau_{f+1}).
\eeq
Equality~\req{gov3333} is the key relation underlying the DFS algorithm. It shows
that the backlog at the end of a frame depends only on the new arrivals during
that frame.

\subsection{Stability Analysis}

In this section, we analyze the stability of the DFS algorithm. For clarity of
exposition, we specialize the arrival model to independent Poisson arrivals.

\begin{description}
\item[(A1)]
For each pair $((i,\ell),(j,k))$, the arrival processes
$\{a_{(i,\ell),(j,k)}(t)\}_{t\ge 1}$ are independent across $(i,\ell,j,k)$ and
i.i.d.\ over time~$t$, with
$a_{(i,\ell),(j,k)}(t)\sim \mathrm{Pois}(\lambda_{(i,\ell),(j,k)})$.
Equivalently, for any integers $s<t$,
\[
\sum_{u=s+1}^{t} a_{(i,\ell),(j,k)}(u)
\sim
\mathrm{Pois}(\lambda_{(i,\ell),(j,k)}(t-s)).
\]
\end{description}

Such an assumption can be extended to more general stochastic processes,
including finite-state Markov, renewal, and autoregressive arrival processes, as
in~\cite{Chang95,Chang2009}.

Our main result of this section is to
establish a stability result for the DFS algorithm.

\bthe{stability}
Consider the DFS algorithm with intra-server traffic balancing.
Let 
\beq{stab0011}
\lambda_{i,j}=\frac{1}{m}\sum_{\ell=1}^m \sum_{k=1}^m  \lambda_{(i,\ell),(j,k)}.
\eeq
Assume that the arrival processes satisfy condition~(A1) and
\bear{stab1111m}
&&\sum_{j=1}^n  \lambda_{i,j} < 1,
\qquad \forall\, i \label{eq:stab1111ma}\\
&&\sum_{i=1}^n  \lambda_{i,j} < 1,
\qquad \forall\, j. \label{eq:stab1111mb}
\eear
Then, for any frame $f\ge 1$,
\beq{cb9900m}
\mathbb{E}[T_f] < \infty .
\eeq
\ethe

\bproof
Fix a frame index $f\ge 1$.  Let
\[
W_{i,j}(f)
\triangleq
\sum_{\ell=1}^m\sum_{k=1}^m
A_{(i,\ell),(j,k)}(\tau_f,\tau_{f+1})
\]
denote the total number of packets that arrive during frame $f$ and are destined
from server $i$ to server $j$.  Define the server-level aggregate arrivals
\[
U_i(f)\triangleq \sum_{j=1}^n W_{i,j}(f),
\qquad
V_j(f)\triangleq \sum_{i=1}^n W_{i,j}(f),
\]
which represent, respectively, the total number of packets arriving during frame
$f$ that originate from server $i$ (to all destinations) and that are destined to
server $j$ (from all sources).

\medskip
\noindent\textbf{Step 1: Frame length under intra-server traffic balancing.}
By Step~(i) of DFS, packets arriving during frame $f$ are served only in frame
$f+1$.  At the beginning of frame $f+1$, DFS applies intra-server traffic
balancing (Step~(iii)) before scheduling the $mn\times mn$ crossbar (Step~(iv)).
The balancing procedure guarantees that, after balancing, the total \emph{outgoing}
traffic load assigned to each of the $m$ NICs in server $i$ is at most
$\lceil U_i(f)/m\rceil$, and the total \emph{incoming} traffic load assigned to
each NIC in server $j$ is at most $\lceil V_j(f)/m\rceil$.
Therefore, the minimum completion (clearance) time for the balanced traffic in
frame $f+1$ satisfies
\begin{equation}
T_{f+1}
=
\max\!\left\{
\max_{1\le i\le n}\left\lceil \frac{U_i(f)}{m}\right\rceil,
\;
\max_{1\le j\le n}\left\lceil \frac{V_j(f)}{m}\right\rceil
\right\}.
\label{eq:Tf_balanced}
\end{equation}
Using $\lceil x\rceil \le x+1$ for all $x\in\mathbb{R}$, we further obtain
\begin{equation}
T_{f+1}
\le
1+
\max\!\left\{
\max_{1\le i\le n}\frac{U_i(f)}{m},
\;
\max_{1\le j\le n}\frac{V_j(f)}{m}
\right\}.
\label{eq:Tf_ceiling_bound}
\end{equation}

\medskip
\noindent\textbf{Step 2: Exponential moment bound.}
Fix $\theta>0$.  From \req{Tf_ceiling_bound} and $\max\{z_1,z_2\}\le z_1+z_2$
for $z_1,z_2\ge 0$, we have
\bear{exp_T_bound}
&&e^{\theta T_{f+1}}\nonumber\\
&&\le
e^{\theta}\,
\max\!\Bigg\{
\max_{1\le i\le n} \exp\!\Big(\tfrac{\theta}{m}U_i(f)\Big),
\nonumber\\
&&\qquad\qquad\qquad
\max_{1\le j\le n} \exp\!\Big(\tfrac{\theta}{m}V_j(f)\Big)
\Bigg\}
\nonumber\\
&&\le
e^{\theta}\,
\left(
\sum_{i=1}^n \exp\!\Big(\tfrac{\theta}{m}U_i(f)\Big)
+
\sum_{j=1}^n \exp\!\Big(\tfrac{\theta}{m}V_j(f)\Big)
\right).
\eear

\medskip
\noindent\textbf{Step 3: Conditional MGFs under Poisson arrivals.}
Under (A1), for each fixed $(i,j)$, $W_{i,j}(f)$ is Poisson with mean
\[
\mathbb{E}[W_{i,j}(f)\mid T_f]
=
T_f\sum_{\ell=1}^m\sum_{k=1}^m \lambda_{(i,\ell),(j,k)}
=
m\,\lambda_{i,j}\,T_f,
\]
where $\lambda_{i,j}$ is defined in \req{stab0011}.  Moreover, for fixed $i$,
$\{W_{i,j}(f)\}_{j=1}^n$ are independent (superposition of independent Poisson
variables), hence $U_i(f)=\sum_{j}W_{i,j}(f)$ is Poisson with mean
\[
\mathbb{E}[U_i(f)\mid T_f]
=
mT_f \sum_{j=1}^n \lambda_{i,j}.
\]
Similarly, for fixed $j$, $V_j(f)$ is Poisson with mean
\[
\mathbb{E}[V_j(f)\mid T_f]
=
mT_f \sum_{i=1}^n \lambda_{i,j}.
\]
Therefore, conditioning on $T_f$ and using the Poisson MGF,
\[
\mathbb{E}\!\left[e^{\eta Z}\right]=\exp\!\big(\mu(e^{\eta}-1)\big)
\quad\text{for }Z\sim\mathrm{Pois}(\mu),
\]
we obtain, almost surely,
\bear{mgf}
&&\sum_{i=1}^n \mathbb{E}\!\left[\exp\!\Big(\tfrac{\theta}{m}U_i(f)\Big)\,\middle|\,T_f\right]\nonumber\\
&&=
\sum_{i=1}^n
\exp\!\Bigg(
mT_f\Big(\sum_{j=1}^n \lambda_{i,j}\Big)\big(e^{\theta/m}-1\big)
\Bigg),
\label{eq:Ui_mgf}
\\
&&\sum_{j=1}^n \mathbb{E}\!\left[\exp\!\Big(\tfrac{\theta}{m}V_j(f)\Big)\,\middle|\,T_f\right]\nonumber\\
&&=
\sum_{j=1}^n
\exp\!\Bigg(
mT_f\Big(\sum_{i=1}^n \lambda_{i,j}\Big)\big(e^{\theta/m}-1\big)
\Bigg).
\label{eq:Vj_mgf}
\eear

Let
\[
\bar\lambda
\triangleq
\max\!\left\{
\max_{1\le i\le n}\sum_{j=1}^n \lambda_{i,j},
\;
\max_{1\le j\le n}\sum_{i=1}^n \lambda_{i,j}
\right\}.
\]
By \req{stab1111ma} and \req{stab1111mb}, we have $\bar\lambda<1$.  Combining
\req{exp_T_bound}--\req{Vj_mgf} and taking expectations yield
\begin{equation}
\mathbb{E}\!\left[e^{\theta T_{f+1}}\right]
\le
2n\,e^{\theta}\,
\mathbb{E}\!\left[
\exp\!\Big(m\bar\lambda\,T_f\,(e^{\theta/m}-1)\Big)
\right].
\label{eq:mgf_rec_raw_balanced}
\end{equation}

\medskip
\noindent\textbf{Step 4: Choose $\theta^*$ and derive a contraction.}
Since $(e^{x}-1)/x$ is strictly increasing for $x>0$ and ranges from $1$ to
$\infty$, and since $(1+\bar\lambda)/(2\bar\lambda)>1$ (because $\bar\lambda<1$),
there exists a unique $\theta^*>0$ such that
\begin{equation}
\frac{e^{\theta^*/m}-1}{\theta^*/m}
=
\frac{1+\bar\lambda}{2\bar\lambda}.
\label{eq:theta_star_balanced}
\end{equation}
Equivalently,
\[
m\bar\lambda\,(e^{\theta^*/m}-1)
=
\theta^*\,\frac{1+\bar\lambda}{2}.
\]
Substituting $\theta=\theta^*$ into \req{mgf_rec_raw_balanced} gives
\begin{equation}
\log \mathbb{E}\!\left[e^{\theta^* T_{f+1}}\right]
\le
\log(2n)+\theta^*
+
\log \mathbb{E}\!\left[
e^{\theta^*\frac{1+\bar\lambda}{2} T_f}
\right].
\label{eq:log_mgf_rec_balanced}
\end{equation}

Let $\phi(\vartheta)\triangleq \log\mathbb{E}[e^{\vartheta T_f}]$, which is convex
in $\vartheta$ (see, e.g., \cite{Chang2000}, Proposition~7.1.8).  With
$\alpha\triangleq (1+\bar\lambda)/2\in(0,1)$, Jensen's inequality for convex
$\phi(\cdot)$ yields
\bearn
&&\log \mathbb{E}\!\left[e^{\alpha\theta^* T_f}\right]
=
\phi(\alpha\theta^*)\\
&&\le
(1-\alpha)\phi(0)+\alpha\phi(\theta^*)
=
\alpha\log \mathbb{E}\!\left[e^{\theta^* T_f}\right].
\eearn
Applying this bound to \req{log_mgf_rec_balanced} gives the contraction
\begin{equation}
\log \mathbb{E}\!\left[e^{\theta^* T_{f+1}}\right]
\le
\log(2n)+\theta^*
+
\frac{1+\bar\lambda}{2}\,
\log \mathbb{E}\!\left[e^{\theta^* T_f}\right].
\label{eq:log_mgf_contraction_balanced}
\end{equation}

\medskip
\noindent\textbf{Step 5: Uniform boundedness of the log-MGF and finiteness of $\mathbb{E}[T_f]$.}
Define $y_f\triangleq \log \mathbb{E}[e^{\theta^* T_f}]$.  Then
\eqref{eq:log_mgf_contraction_balanced} implies
\[
y_{f+1}
\le
\big(\log(2n)+\theta^*\big)
+
\alpha y_f,
\qquad
\alpha=\frac{1+\bar\lambda}{2}\in(0,1).
\]
Since $T_1=1$, we have $y_1=\theta^*<\infty$, and the above recursion implies that
$\sup_f y_f<\infty$ (e.g., by iterating the inequality).  In particular, for all
$f\ge 1$,
\bearn
&&y_f \le \max\!\left\{ \theta^*,\,
\frac{\log(2n)+\theta^*}{1-\alpha}
\right\} \\
&&=
\max\!\left\{
\theta^*,
\,
\frac{2(\log(2n)+\theta^*)}{1-\bar\lambda}
\right\}
<\infty.
\eearn
Finally, Jensen's inequality (convexity of $e^{\theta^* x}$) yields
\[
\mathbb{E}[T_f]
\le
\frac{1}{\theta^*}\log \mathbb{E}\!\left[e^{\theta^* T_f}\right]
=
\frac{y_f}{\theta^*}
<\infty,
\qquad \forall f\ge 1,
\]
which proves \req{cb9900m} and completes the proof of \rthe{stability}.
\eproof

\bsec{Simulation}{simulation1}

In this section, we present simulation results to evaluate the performance of
the proposed DFS algorithm with hierarchical Birkhoff--von Neumann
decomposition.

\subsection{Traffic Models}
\label{sec:traffic_models}

In this subsection, we propose two server/GPU-level traffic models for
simulation studies.
The system consists of an $mn\times mn$ input-buffered
crossbar switch, where $n=8$ servers and each server contains $m=2$ GPUs,
resulting in a $16\times16$ switch fabric. As in our analysis, we assume Poisson arrivals as in (A1).
Both models are specified via the \emph{arrival rate matrix}
\[
r_{(i,\ell),(j,k)} \;\triangleq\; \mathbb{E}\big[a_{(i,\ell),(j,k)}(t)\big],
\]
where $(i,\ell)$ denotes the $\ell$-th GPU (and NIC) in server $i$, and
$a_{(i,\ell),(j,k)}(t)$ is the number of packets that arrive at time slot $t$ and
are destined from $(i,\ell)$ to $(j,k)$.

Throughout, we focus on \emph{inter-server} traffic and set all intra-server
traffic rates to zero, i.e.,
$r_{(i,\ell),(i,k)}=0$ for all $i$ and all $\ell,k\in\{1,\ldots,m\}$.
(Equivalently, the diagonal $m\times m$ blocks are zero.)

\noindent{\bf Model U: Uniform GPU-to-GPU Traffic}

In the \emph{uniform} traffic model, all inter-server GPU-to-GPU rates are
identical:
\bear{uniform_rate}
&&r^{\mathrm{U}}_{(i,\ell),(j,k)}
\;=\;
\begin{cases}
r_0, & i\neq j,\\
0, & i=j,
\end{cases}
\nonumber \\
&&\qquad
\forall\, i,j\in\{1,\ldots,n\},\ \forall\, \ell,k\in\{1,\ldots,m\},
\eear
where $r_0>0$ is chosen to achieve a prescribed offered load.

Under Model~U, the server-level aggregated rates
\[
\lambda^{\mathrm{U}}_{i,j}
\;\triangleq\;
\frac{1}{m}\sum_{\ell=1}^m\sum_{k=1}^m r^{\mathrm{U}}_{(i,\ell),(j,k)}
\]
satisfy $\lambda^{\mathrm{U}}_{i,j}=m r_0$ for $i\neq j$, and $0$ otherwise.
Thus, the aggregate traffic is also uniform across server pairs.

\noindent{\bf Model NU: Non-Uniform (Server-Localized Hotspot) Traffic}

In the \emph{non-uniform} traffic model, we intentionally create a per-server
hotspot by concentrating all inter-server traffic of each server pair $(i,j)$
onto a single GPU pair, while preserving the server-level aggregate rates of
Model~U.

Specifically, define
\begin{equation}
\tilde r^{\mathrm{NU}}_{(i,\ell),(j,k)}
\;=\;
\begin{cases}
\displaystyle m^2 r_0,
& (i\neq j)\ \text{and}\ (\ell,k)=(1,1),\\[2mm]
0, & \text{otherwise}.
\end{cases}
\label{eq:nonuniform_rate_def}
\end{equation}

By construction, Model~NU preserves the server-level aggregate rates:
\begin{equation}
\frac{1}{m}\sum_{\ell=1}^m\sum_{k=1}^m \tilde r^{\mathrm{NU}}_{(i,\ell),(j,k)}
\;=\;
\frac{1}{m}\sum_{\ell=1}^m\sum_{k=1}^m r^{\mathrm{U}}_{(i,\ell),(j,k)}
\;=\;
mr_0,
\label{eq:aggregate_preserved}
\end{equation}
for all $i, j$.
Thus, Models~U and~NU have the same inter-server \emph{aggregate} load, but Model~NU
exhibits extreme \emph{micro-level} non-uniformity across GPUs/NICs within a
server.

\begin{figure}[t]
  \centering
  \subfloat[Model U]{\includegraphics[width=0.49\linewidth]{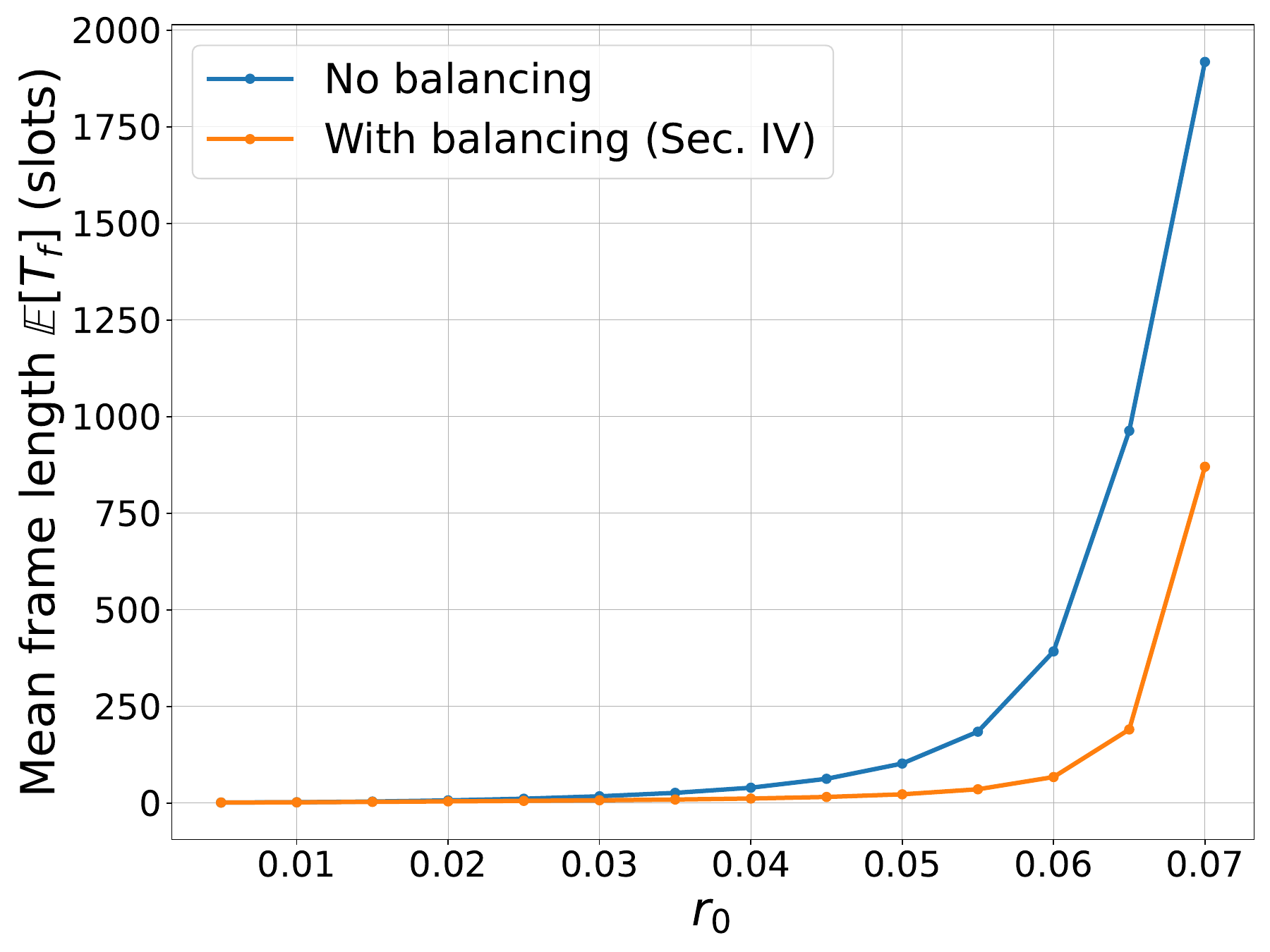}\label{fig:mean_frame_U}}
  \hfill
  \subfloat[Model NU]{\includegraphics[width=0.49\linewidth]{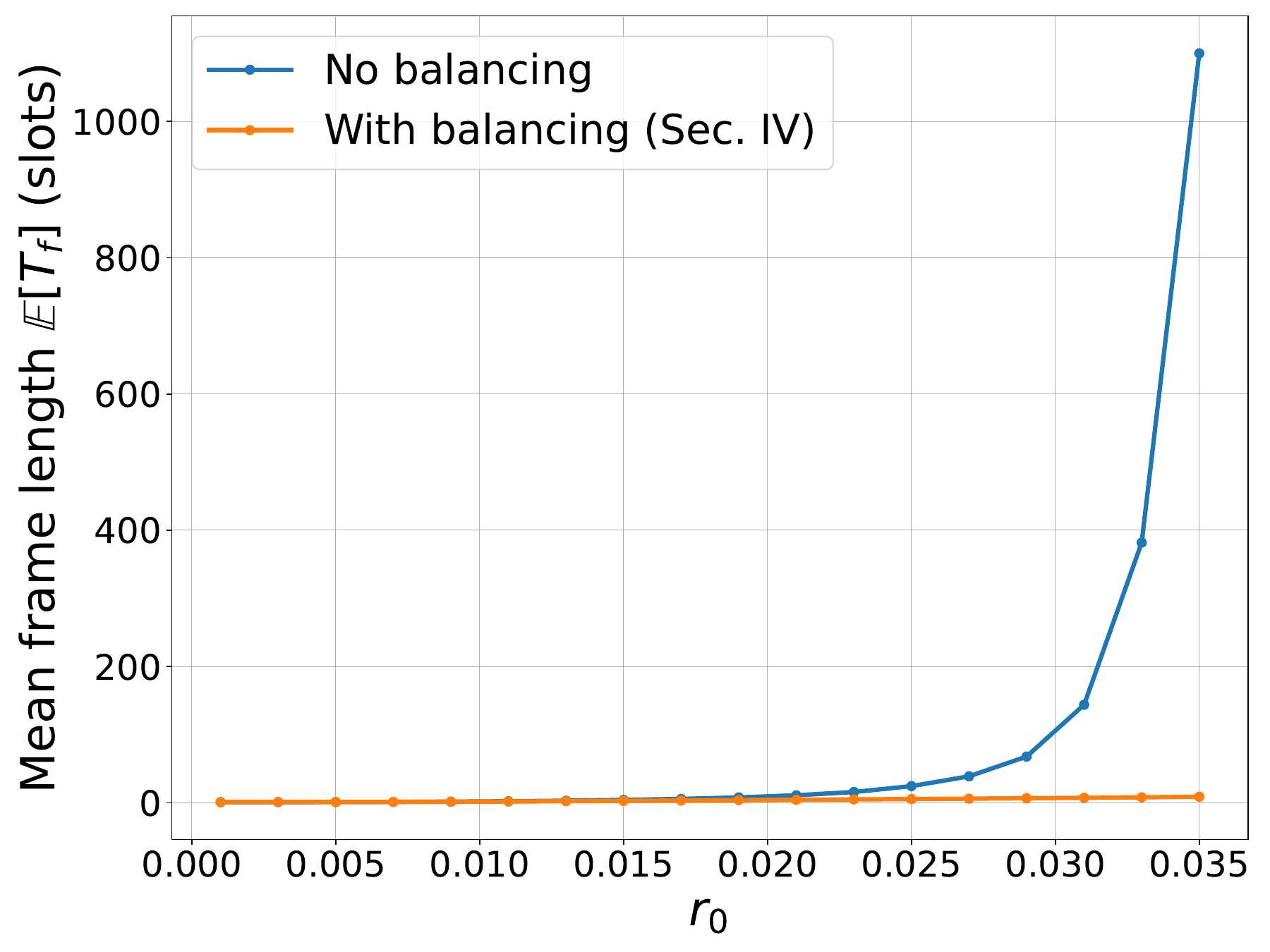}\label{fig:mean_frame_NU}}
  \caption{Mean frame length $\mathbb{E}[T_f]$ versus $r_0$ with $(n,m)=(8,2)$ under (a) Model~U and (b) Model~NU. Curves compare DFS with hierarchical BvN decomposition with and without intra-server balancing (Sec.~\ref{sec:balancing}). Each point is obtained from a fixed slot-horizon simulation with a warm-up period removed from statistics.}
  \label{fig:mean_frame_models}
\end{figure}

\subsection{Schemes Compared}
\label{sec:schemes_compared}

We compare two variants of DFS with hierarchical BvN decomposition:
\begin{itemize}
\item \emph{No intra-server balancing:} DFS with hierarchical BvN decomposition applied directly to the per-GPU backlog matrix.
\item \emph{With intra-server balancing (Sec.~\ref{sec:balancing}):} DFS with intra-server traffic balancing applied at each frame boundary, followed by the same hierarchical BvN decomposition.
\end{itemize}
Both variants use the same DFS mechanism; the only difference is whether intra-server balancing is enabled.

\subsection{Simulation Settings and Metric}
\label{sec:sim_settings}

All results are obtained using a \emph{fixed slot-horizon} simulation to ensure fair
comparisons between schemes that may induce different frame sizes.
Unless otherwise stated, each run simulates $10^5$ time slots, and the first
$10^4$ slots are treated as warm-up and excluded from statistics.
We sweep $r_0$ over $[0.005,0.07]$ for Model~U and over $[0.001,0.035]$ for Model~NU.

We report the \emph{mean frame length} $\mathbb{E}[T_f]$, computed by averaging $T_f$
over all frames observed within the simulated horizon after the warm-up period.

\subsection{Results: Mean Frame Length Versus $r_0$}
\label{sec:sim_results}

Fig.~\ref{fig:mean_frame_models} shows the mean frame length $\mathbb{E}[T_f]$ as a function of $r_0$
for Models~U and~NU. For reference, under Model~U per-port capacity condition corresponds to $r_0<1/((n-1)m)=1/14$, while under Model~NU without balancing it corresponds to $r_0<1/((n-1)m^2)=1/28$.

\subsubsection{Model U (uniform micro-level traffic)}
In Fig.~\ref{fig:mean_frame_U}, both schemes exhibit increasing mean frame length as $r_0$ grows,
with a rapid rise as $r_0$ approaches $1/14$. 
Although Model~U is uniform at the GPU-to-GPU rate level, enabling intra-server balancing still
reduces $\mathbb{E}[T_f]$ over the entire range of $r_0$.
This suggests that balancing can reduce the mean frame length by smoothing within-block load variations caused by arrival randomness, even when the underlying rate matrix is uniform.

\subsubsection{Model NU (server-localized hotspot)}
In Fig.~\ref{fig:mean_frame_NU}, the no-balancing scheme suffers a much sharper increase in the mean frame length
$\mathbb{E}[T_f]$ as $r_0$ increases, which is consistent with the severe micro-level concentration of traffic onto the
first GPU pair $(i,1)\!\to\!(j,1)$ for each inter-server pair $(i,j)$. In contrast, enabling intra-server balancing
(Sec.~\ref{sec:balancing}) substantially reduces $\mathbb{E}[T_f]$ over the entire sweep range and delays the onset of very large frames.
Overall, Model~NU clearly separates the two schemes: without balancing, hotspot-induced per-GPU stragglers inflate the
frame length, whereas intra-server balancing redistributes traffic within each $m\times m$ server-pair block and
significantly improves frame-size behavior.

\begin{figure}[t]
  \centering
  \includegraphics[width=0.80\linewidth]{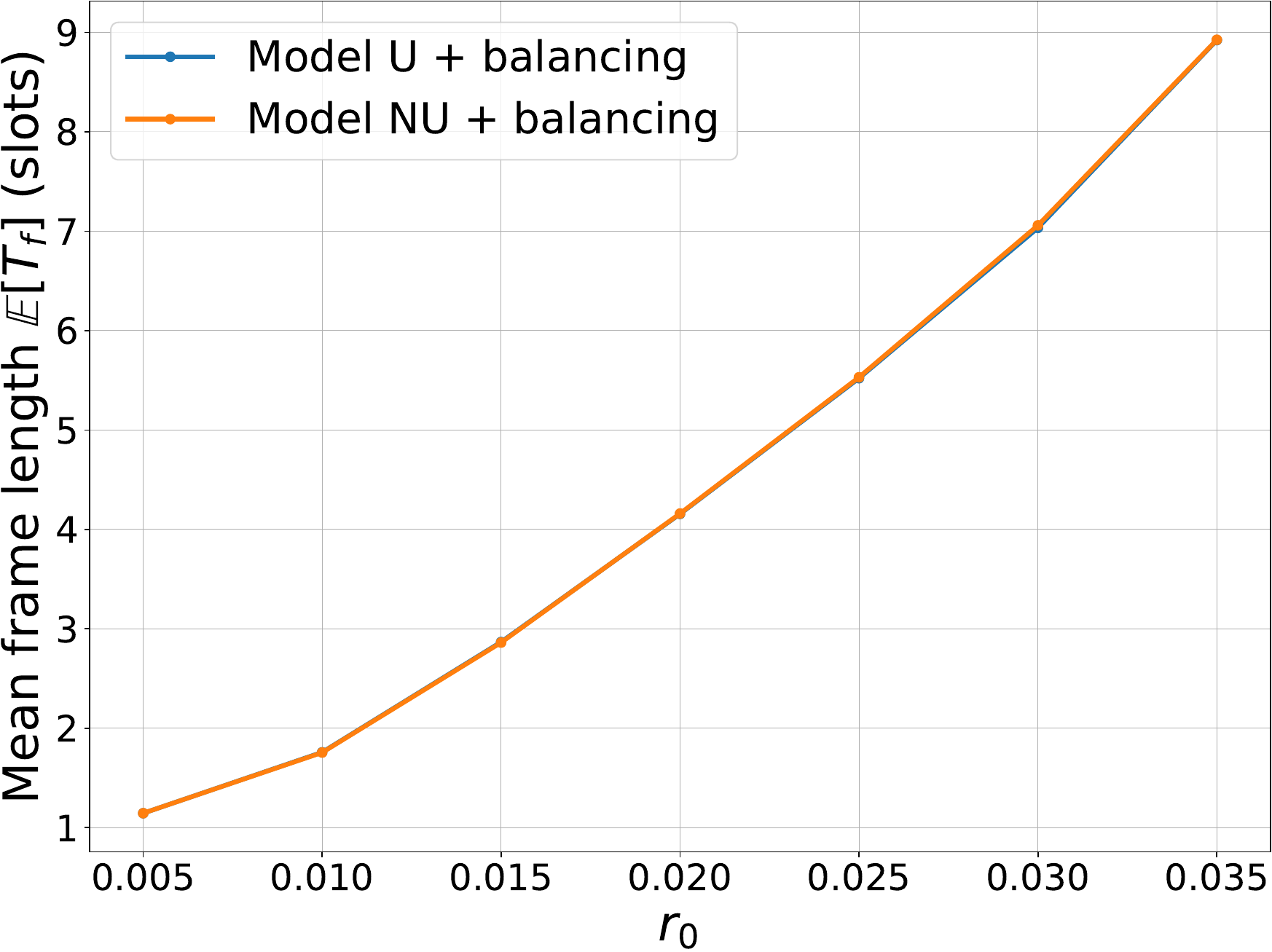}
  \caption{Mean frame length $\mathbb{E}[T_f]$ versus $r_0$ with $(n,m)=(8,2)$ under intra-server balancing (Sec.~\ref{sec:balancing}). The plot overlays Model~U and Model~NU over the common sweep range of $r_0$, showing that the two curves nearly overlap.}
  \label{fig:U_vs_NU_bal}
\end{figure}

Moreover, Fig.~\ref{fig:U_vs_NU_bal} overlays the balanced cases for Models~U and~NU over the common range of $r_0$.
The two curves nearly overlap, indicating that intra-server balancing effectively removes the micro-level non-uniformity
in Model~NU and yields frame-length behavior comparable to that of the uniform Model~U.

\subsubsection{Summary}
Across both traffic models, intra-server balancing consistently reduces the mean frame length $\mathbb{E}[T_f]$.
The improvement is especially pronounced under Model~NU, where traffic is highly concentrated within each server pair.
These results support the role of intra-server balancing (Sec.~\ref{sec:balancing}) as a practical mechanism to enhance robustness of DFS
with hierarchical BvN decomposition under heterogeneous GPU-level traffic patterns.

\section{Conclusion}
\label{sec:conclusion}

This paper develops an online scheduling framework for all-to-all GPU communication in two-tier clusters with fast intra-server links and a slower inter-server fabric.
Building on the dynamic frame sizing (DFS) principle, we propose a \emph{dynamic hierarchical} Birkhoff--von Neumann (BvN) decomposition that exploits the $n$-server structure to avoid the prohibitive cost of directly decomposing an $mn\times mn$ GPU-level matrix.
At each frame boundary, we additionally perform a simple \emph{intra-server traffic balancing} step using only local unit-transfer operations, reshaping each $m\times m$ server-pair block to reduce micro-level (GPU/NIC-level) skew while preserving the aggregate server-to-server demand.

On the theoretical side, we (i) characterize a completion-time lower bound induced by per-port capacities, (ii) show that intra-server balancing does not increase this bound and can reduce it by equalizing per-GPU outgoing/incoming loads within each server pair, and (iii) establish a DFS stability guarantee under admissible Poisson arrivals expressed in terms of server-level aggregate loads.
On the empirical side, simulations under both a uniform micro-level model (Model~U) and a server-localized hotspot model (Model~NU) show that balancing consistently reduces the mean frame length $\mathbb{E}[T_f]$, with particularly large gains under hotspot traffic; after balancing, the frame-length behavior under Model~NU closely matches that under Model~U, indicating effective removal of micro-level non-uniformity.

Overall, the results suggest that combining hierarchical decomposition with lightweight intra-server balancing yields a practical and robust approach to scheduling large-scale all-to-all transfers under per-port matching constraints.
Future work includes extending the framework to more realistic fabrics and traffic primitives, such as multi-hop interconnects, heterogeneous link rates, variable packet sizes, and additional topology constraints, as well as developing online/learning-based estimation of the aggregated server-level traffic matrix to further reduce control overhead.




\begin{IEEEbiography}[{\includegraphics[width=1in,height=1.25in,clip,keepaspectratio]{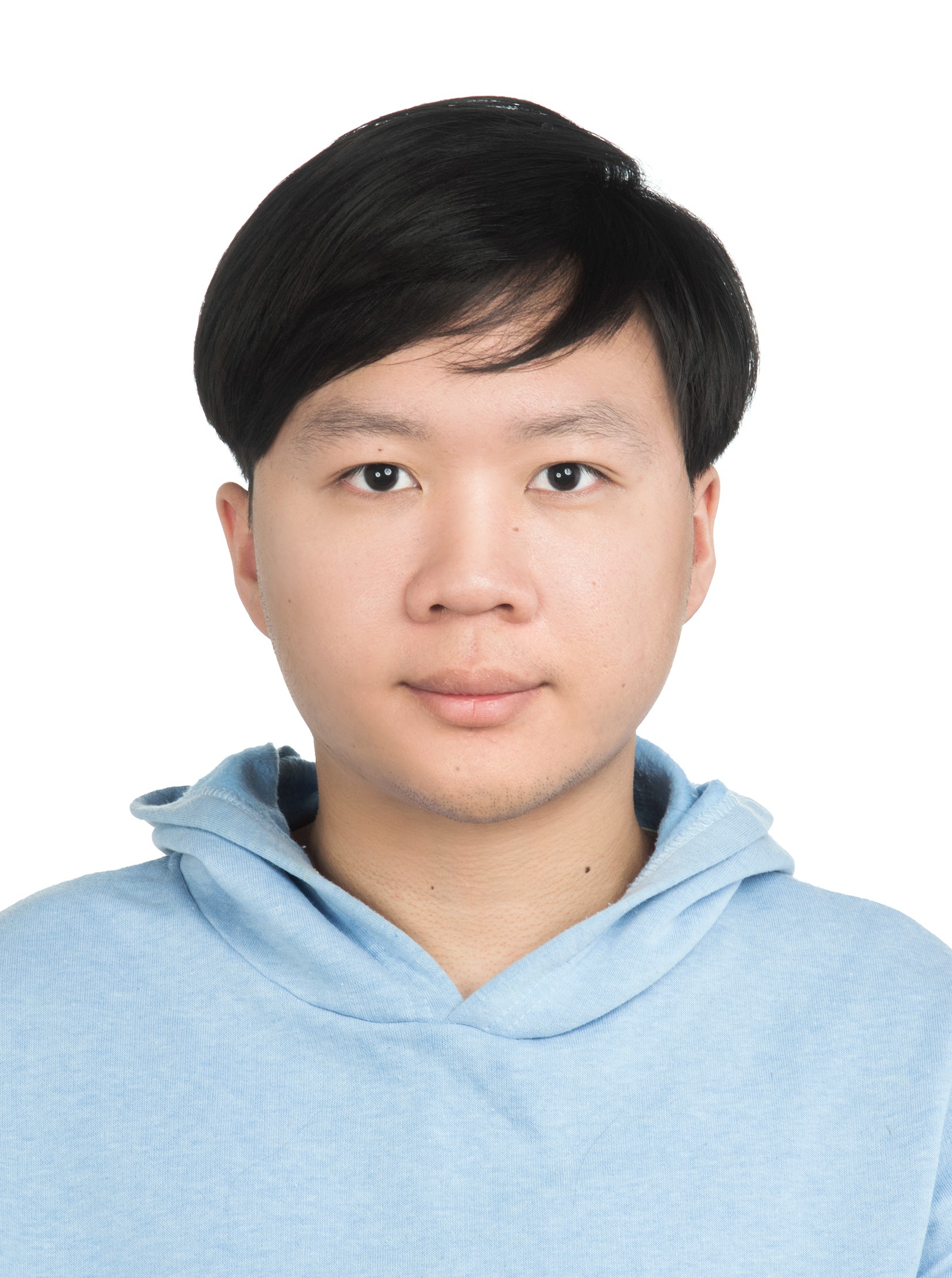}}]
{Yen-Chieh Wu} received the B.S. degree in electrical engineering in 2024 from National Tsing Hua University, Hsinchu, Taiwan. He is currently pursuing the M.S. degree in the Institute of Communications Engineering, National Tsing Hua University, Hsinchu, Taiwan. His research interests include communication network scheduling and large language model (LLM) agents.
\end{IEEEbiography}

\begin{IEEEbiography}[{\includegraphics[width=1in,height=1.25in,clip,keepaspectratio]{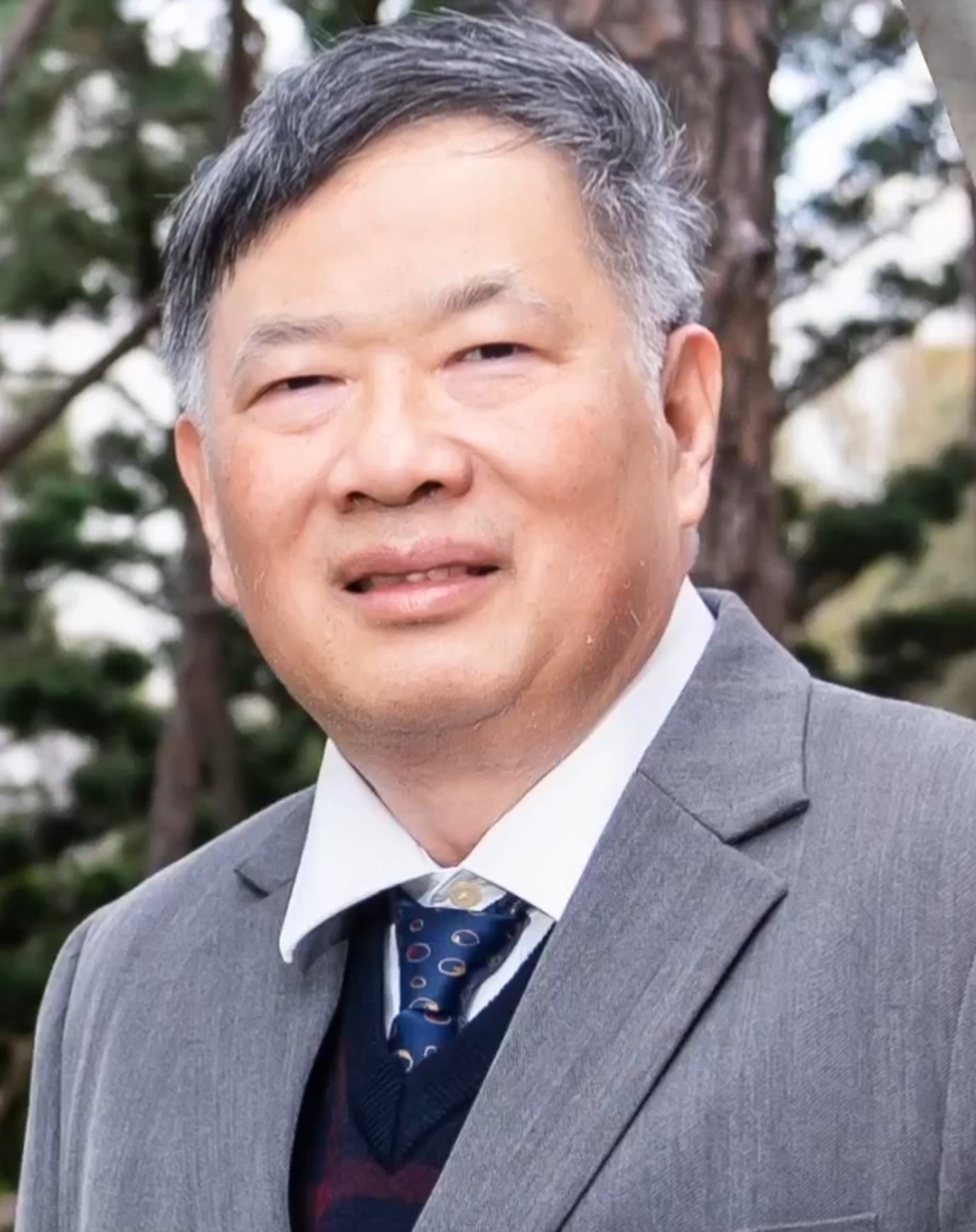}}]
{Cheng-Shang Chang}(S'85-M'86-M'89-SM'93-F'04) received the B.S. degree from National Taiwan University, Taipei, Taiwan, in 1983, and the M.S. and Ph.D. degrees from Columbia University, New York, NY, USA, in 1986 and 1989, respectively, all in electrical engineering.
From 1989 to 1993, he was employed as a Research Staff Member with the IBM Thomas J. Watson Research Center, Yorktown Heights, NY, USA. Since 1993, he has been with the Department of Electrical Engineering, National Tsing Hua University, Taiwan, where he is a Tsing Hua Distinguished Chair Professor. Dr. Chang served as an Editor for Operations Research from 1992 to 1999, an Editor for the {\em IEEE/ACM TRANSACTIONS ON NETWORKING} from 2007 to 2009, and an Editor for the {\em IEEE TRANSACTIONS ON NETWORK SCIENCE AND ENGINEERING} from 2014 to 2017. He is currently serving as an Editor-at-Large for the {\em IEEE/ACM TRANSACTIONS ON NETWORKING}. He received an IBM Outstanding Innovation Award in 1992, an IBM Faculty Partnership Award in 2001, and Outstanding Research Awards from the National Science Council, Taiwan, in 1998, 2000, and 2002, respectively. He received the Merit NSC Research Fellow Award from the National Science Council, R.O.C. in 2011. He also received the Academic Award in 2011 and the National Chair Professorship in 2017 and 2023 from the Ministry of Education, R.O.C. He is the recipient of the 2017 IEEE INFOCOM Achievement Award.
\end{IEEEbiography}

\begin{IEEEbiography}[{\includegraphics[width=1in,height=1.25in,clip,keepaspectratio]{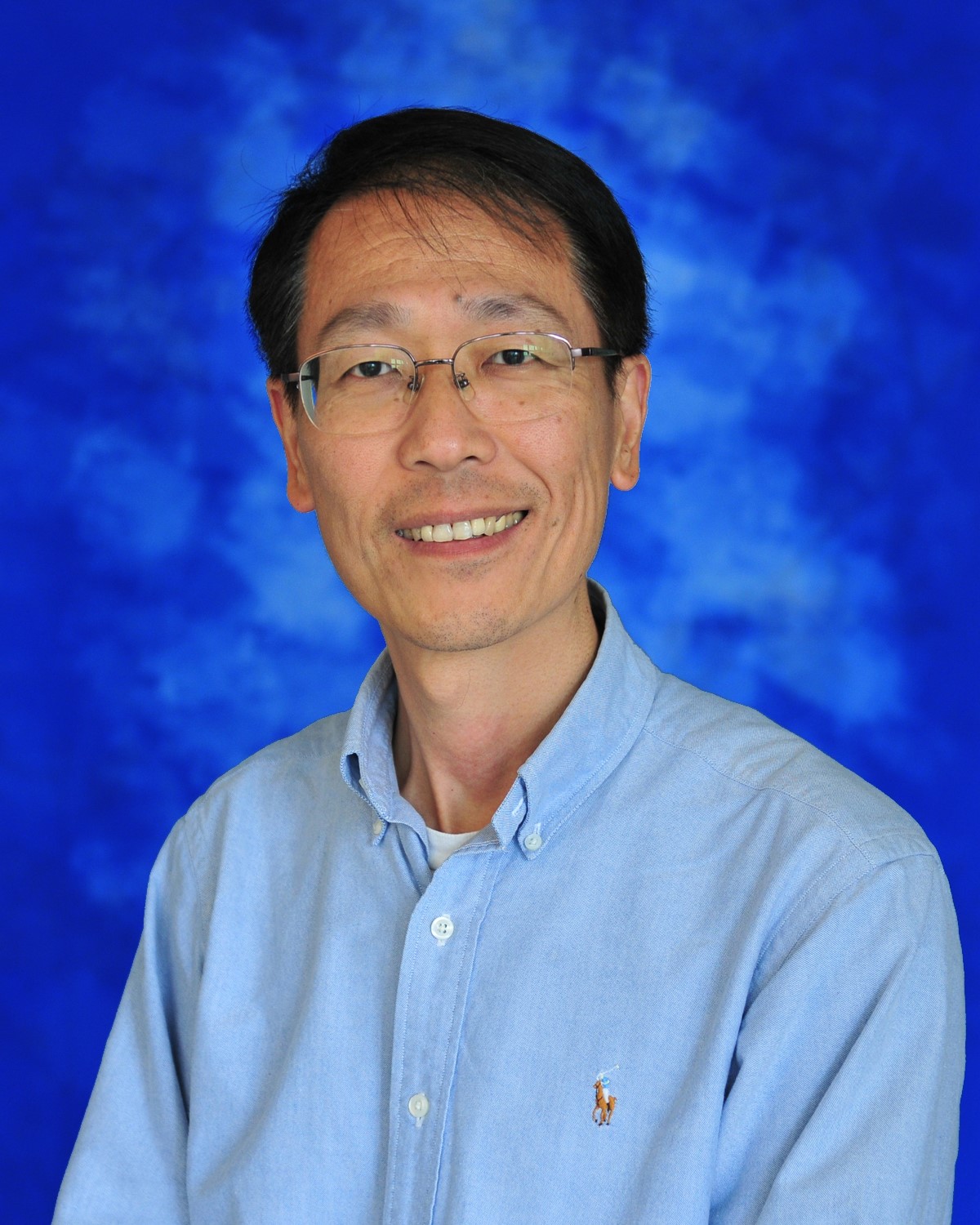}}]
{Duan-Shin Lee}(S'89-M'90-SM'98) received the B.S. degree from National Tsing Hua University, Taiwan, in 1983, and the MS and Ph.D. degrees from Columbia University, New York, in 1987 and 1990, all in electrical engineering. He worked as a research staff member at the C\&C Research Laboratory of NEC USA, Inc. in Princeton, New Jersey from 1990 to 1998. He joined the Department of Computer Science of National Tsing Hua University in Hsinchu, Taiwan, in 1998. Since August 2003, he has been a professor. He received a best paper award from the Y.Z. Hsu Foundation in 2006. He served as an editor for the Journal of Information Science and Engineering between 2013 and 2015.  He is currently an editor for Performance Evaluation. Dr. Lee's current research interests are network science, game theory, machine learning and high-speed networks. He is a senior IEEE member.
\end{IEEEbiography}

\begin{IEEEbiography}[{\includegraphics[width=1in,height=1.25in,clip,keepaspectratio]{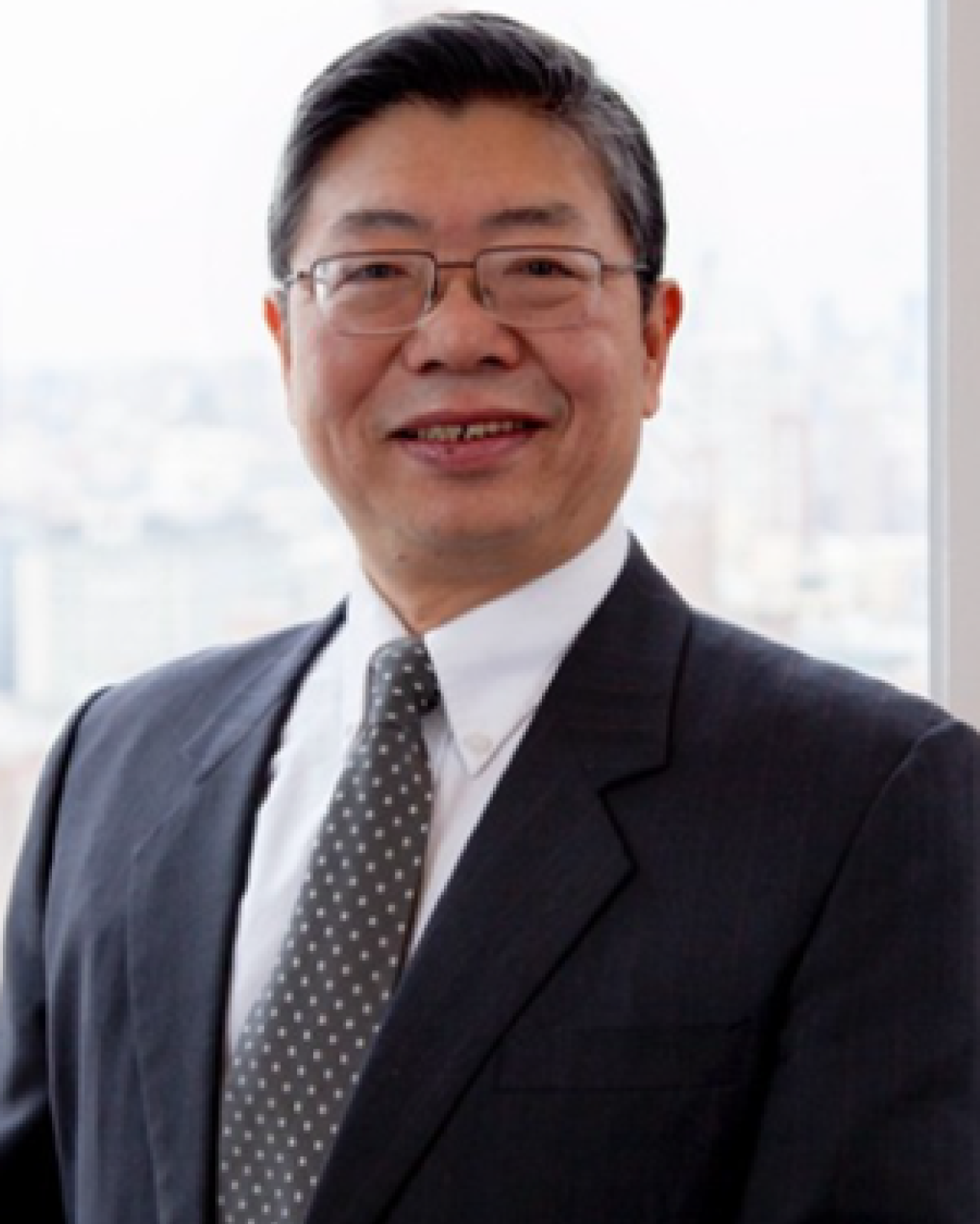}}]
{H.~Jonathan Chao}(Life Fellow, IEEE) received the B.S. and M.S. degrees in electrical engineering
from National Chiao Tung University, Taiwan, in 1977 and 1980, respectively, and the Ph.D. degree
in electrical engineering from The Ohio State University, Columbus, OH, USA, in 1985.
In January 1992, he joined New York University (NYU), where he is currently a Professor of electrical and computer
engineering (ECE). He was the Head of the Department from 2004 to 2014. He is also the Director of the
High-Speed Networking Laboratory, where he is conducting research in the areas of AI datacenter designs to accelerate
AI training and inference, dynamic multi-path load balancing, routing and scheduling, high-speed packet
processing/switching/routing, software-defined networking, network security, and network on chip.
From 2000 to 2001,
he was the Co-Founder and the CTO of Coree Networks, Tinton Falls,
NJ, USA. From 1985 to 1992, he was a member of Technical Staff at
Bellcore, Piscataway, NJ, USA, where he was involved in transport and
switching system architecture designs and application-specified integrated
circuit implementations, such as the world's first SONET-like framer chip,
ATM layer chip, sequencer chip (the first chip handling packet scheduling),
and ATM switch chip. He has co-authored three networking books, \textit{Broadband
Packet Switching Technologies—A Practical Guide to ATM Switches and IP
Routers} (New York: Wiley, 2001), \textit{Quality of Service Control in High-Speed
Networks} (New York: Wiley, 2001), and \textit{High-Performance Switches and
Routers} (New York: Wiley, 2007). He holds 63 patents and has published
more than 300 journals and conference papers. He is a Fellow of the National
Academy of Inventors. He was a recipient of the Bellcore Excellence Award
in 1987. He was a co-recipient of the 2001 Best Paper Award from {\em IEEE
TRANSACTION ON CIRCUITS AND SYSTEMS FOR VIDEO TECHNOLOGY}.
\end{IEEEbiography}

\end{document}